\documentclass[11pt, letterpaper]{article}

\usepackage{graphicx}
\usepackage{amsmath}
\usepackage{amssymb}
\usepackage{setspace}
\usepackage[letterpaper,left=1in,right=1in,top=1in,bottom=1in]{geometry}
\usepackage{color,soul}

\usepackage{subfigure}
\usepackage{float}
\usepackage{accents}
\usepackage{multirow}

\title{\Large \bf Sensitivity-based dynamic performance assessment for model predictive control with Gaussian noise}

\author{
    \centerline{\normalsize Jianbang Liu$^{a}$, Song Bo$^{b}$, Benjamin Decardi-Nelson$^{b}$, Jinfeng Liu$^{b,}$\thanks{Corresponding author: J. Liu. Tel: +1-780-492-1317. Fax: +1-780-492-2881. Email: jinfeng@ualberta.ca.}, Jingtao Hu$^{c}$, Tao Zou$^{d}$}\vspace{5mm}\\
\centerline{\small $^{a}$School of Electrical and Information Engineering, Jiangsu University,}\\
\centerline{\small Zhenjiang 212013, China}\\
\centerline{\small $^{b}$Department of Chemical \& Materials Engineering, University of Alberta,}\\
\centerline{\small Edmonton, AB T6G 1H9, Canada}\\
\centerline{\small $^{c}$Shenyang Institute of Automation, Chinese Academy of Sciences,}\\
\centerline{\small Shenyang 110016, China}\\
\centerline{\small $^{d}$School of Mechanical and Electrical Engineering, Guangzhou University,}\\
\centerline{\small Guangzhou 510006, China}
}

\allowdisplaybreaks

\begin{document}

\date{}
\maketitle
\setstretch{1.35}

\begin{abstract}
Economic model predictive control and tracking model predictive control are two popular advanced process control strategies used in various of fields. Nevertheless, which one should be chosen to achieve better performance in the presence of noise is uncertain when designing a control system. To this end, a sensitivity-based performance assessment approach is proposed to pre-evaluate the dynamic economic and tracking performance of them in this work. First, their controller gains around the optimal steady state are evaluated by calculating the sensitivities of corresponding constrained dynamic programming problems. Second, the controller gains are substituted into control loops to derive the propagation of process and measurement noise. Subsequently, the Taylor expansion is introduced to simplify the calculation of variance and mean of each variable. Finally, the tracking and economic performance surfaces are plotted and the performance indices are precisely calculated through integrating the objective functions and the probability density functions. Moreover, boundary moving (i.e., back off) and target moving can be pre-configured to guarantee the stability of controlled processes using the proposed approach. Extensive simulations under different cases illustrate the proposed approach can provide useful guidance on performance assessment and controller design.
\end{abstract}

\noindent{\bf Keywords:} Dynamic performance assessment, sensitivity analysis, controller gain, model predictive control, Gaussian noise.


\section{Introduction}\label{sec:1}
Over the past decade, economic model predictive control (EMPC), as a generalized form of model predictive control (MPC), has received significant attention \cite{angeli2011average,rawlings2012fundamentals,heidarinejad2012economic,ellis2014tutorial,ellis2017economic}. It has been shown that the asymptotic average performance of an EMPC is never worse than the optimal steady-state operation in the nominal case \cite{angeli2011average}. In \cite{liu2016economic}, an economic model predictive controller with extended horizon was proposed and it was shown that in the nominal condition the transient or asymptotic average performance of EMPC is no worse than the traditional tracking MPC if the prediction horizon is sufficiently large. However, in the presence of model-plant mismatch (e.g., measurement and process noise, process disturbance, parameter uncertainty), whether EMPC can achieve better economic performance than the traditional tracking MPC is uncertain as demonstrated in the applications of EMPC to a post-combustion carbon capture plant \cite{decardi2018improving}, a wind energy conversion system \cite{cui2018comparative}, and a coal-fired boiler-turbine system \cite{zhang2020zone}. There is a need of control performance assessment methods for EMPC in the presence of model-plant mismatch.

In the context of tracking MPC, many studies on performance assessment have been conducted to determine tracking MPC's ideal and reachable performance and to guide the design and implementation of MPC \cite{huang1999performance,jelali2006overview,bauer2008economic,jelali2012control,botelho2016perspectives}. Majority of these methods are based on process operation data. For example, in \cite{wang2006multirate}, Wang and co-workers proposed a data-driven subspace approach for control performance assessment of multi-rate control systems based on minimum variance control. In \cite{lu2016performance}, the idea was extended and applied to the cross-directional control of paper machines. In \cite{xu2007performance}, Xu and co-workers developed an economic performance assessment (EPA) approach and provided constraint/variance tuning guidelines using routine process operation data plus the process steady-state gain matrix. In \cite{wei2007multivariate}, a multivariate economic performance assessment approach was developed and applied to a MPC controlled electric arc furnace. In \cite{lee2008sensitivity}, practical and selective tuning guidelines for MPC was proposed based on sensitivity analysis of process variables. However, the above performance assessment approaches heavily rely on actual process operation data, which are not unavailable before the deployment of a control system.

Based on the above considerations, it is clear that there is a lack of methods that can characterize the control performance of EMPC and tracking MPC in the presence of model-plant mismatch before their actual deployment. Such a method can be very useful in determining which predictive control method (tracking MPC or EMPC) is more appropriate for a specific process. Motivated by this, in this work, we propose a method to evaluate the performance of EMPC and tracking MPC in the presence of Gaussian process and measurement noise.

The remainder of this paper is organized as follows: Section \ref{sec:2} gives a brief introduction of preliminaries, Section \ref{sec:3} illustrates the motivations of this work, Section \ref{sec:4} introduces the implementation of the proposed sensitivity-based dynamic performance assessment approach, Section \ref{sec:5} applies the proposed approach into a chemical process to demonstrate its effectiveness, Section \ref{sec:6} presents the conclusions and future directions.

\section{Preliminaries}\label{sec:2}
\subsection{System description and problem formulation}\label{sec:2.1}
In the work, we consider a general class of discrete time-invariant nonlinear systems described as follows:
\begin{subequations}\label{eq:2.1.1}
  \begin{align}
    {\boldsymbol{x}}(k + 1) &= F({\boldsymbol{x}}(k),{\boldsymbol{u}}(k)) + {\boldsymbol w}(k)\label{eq:2.1.1a}\\
    {\boldsymbol{y}}(k) &= H({\boldsymbol{x}}(k)) + {\boldsymbol v}(k) \label{eq:2.1.1b}
  \end{align}
\end{subequations}
where ${\boldsymbol{x}}(k) \in R^n$, ${\boldsymbol{u}}(k) \in R^m$, and ${\boldsymbol{y}}(k) \in R^r$ denote the state, input, and output vectors at time $k$, respectively; ${\boldsymbol{w}}(k) \in R^n$ and ${\boldsymbol{v}}(k) \in R^r$ denote the process and measurement noise vectors at time $k$, respectively; $F(\cdot)$ and $H(\cdot)$ denote the state and output functions, respectively.

In the work, we focus on characterizing the control performance of economic MPC and tracking MPC in the presence of Gaussian noise. The results can be useful when we need to choose between economic MPC and tracking MPC for a specific control application. The results may also be used to fine tune the control target (operating point) or constraints for improved control performance.

\subsection{Formulation of tracking MPC}\label{sec:2.2}
Conventional tracking MPC minimizes a quadratic cost to track an optimal steady state. In the work, we consider tracking MPC that can be formulated as follows:
\begin{subequations}\label{eq:2.2.1}
  \begin{align}
      \min\limits_{{\hat{\boldsymbol u}(\cdot)},{\hat{\boldsymbol x}(\cdot)}} &J(k) = \sum\limits_{i = k}^{k + N - 1} {\left\| {{\hat{\boldsymbol x}}(i) - {{\boldsymbol{x}}_{\rm{s}}}} \right\|_{\boldsymbol{Q}}^2}  + \sum\limits_{i = k}^{k + N - 1} {\left\| {{\hat{\boldsymbol u}}(i) - {{\boldsymbol{u}}_{\rm{s}}}} \right\|_{\boldsymbol{R}}^2}  \label{eq:2.2.1a}\\
      {\text{s.t.~}}&{\hat{\boldsymbol x}}(i + 1) = F({\hat{\boldsymbol x}}(i),{\hat{\boldsymbol u}}(i)),~\forall i=k,\cdots,k+N-1 \label{eq:2.2.1b}\\
    &{ \hat{\boldsymbol u}}(i) \in {\boldsymbol U},~\forall i=k,\cdots,k+N-1 \label{eq:2.2.1c}\\
    &{ \hat{\boldsymbol x}}(k) = {\boldsymbol x}(k),~{ \hat{\boldsymbol x}}(i) \in {\boldsymbol X},~\forall i=k,\cdots,k+N-1 \label{eq:2.2.1d}
\end{align}
\end{subequations}
where $N$ denotes the prediction horizon; ${\hat{\boldsymbol x}}(\cdot)$ and ${\hat{\boldsymbol u}}(\cdot)$ denote the predicted state and input trajectories over the prediction horizon, respectively; ${\boldsymbol x}(k)$ is the actual system state at instant $k$; ${\boldsymbol x}_{\rm s}$ and ${\boldsymbol u}_{\rm s}$ are the given steady-state targets to be tracked; $\boldsymbol X$ and $\boldsymbol U$ denote the known state and input constraint sets. In (\ref{eq:2.2.1}), (\ref{eq:2.2.1a}) is the cost function of the tracking MPC, in which $\boldsymbol Q$ and $\boldsymbol R$ are positive definite weighting matrices. (\ref{eq:2.2.1b}) is the system model while (\ref{eq:2.2.1c}) and (\ref{eq:2.2.1d}) being the known constraints on states and inputs.

\subsection{Formulation of EMPC}\label{sec:2.3}
EMPC optimizes a general cost function instead of a quadratic cost function. In the work, we consider EMPC that can be formulated as follows:
\begin{subequations}\label{eq:2.3.1}
  \begin{align}
      \min\limits_{{\hat{\boldsymbol u}(\cdot)},{\hat{\boldsymbol x}(\cdot)}} &J(k) = \sum\limits_{i = k}^{k + N - 1} {E({\hat{\boldsymbol x}}(i),{\hat{\boldsymbol u}}(i))} \label{eq:2.3.1a}\\
      {\text{s.t.~}}&{\hat{\boldsymbol x}}(i + 1) = F({\hat{\boldsymbol x}}(i),{\hat{\boldsymbol u}}(i)),~\forall i=k,\cdots,k+N-1 \label{eq:2.3.1b}\\
    &{ \hat{\boldsymbol u}}(i) \in {\boldsymbol U},~\forall i=k,\cdots,k+N-1 \label{eq:2.3.1c}\\
    &{ \hat{\boldsymbol x}}(k) = {\boldsymbol x}(k),~{ \hat{\boldsymbol x}}(i) \in {\boldsymbol X},~\forall i=k,\cdots,k+N-1 \label{eq:2.3.1d}
\end{align}
\end{subequations}
where (\ref{eq:2.3.1a}) is the cost function of the EMPC to minimize and $E(\cdot)$ denotes the economic stage cost.

\subsection{Performance indices}\label{sec:2.4}
In this work, two different performance indices are used: the average tracking performance index $\bar J_{\rm tp}$ and the average economic performance index $\bar J_{\rm ep}$ \cite{bauer2008economic,ellis2014tutorial}:
\begin{subequations}\label{eq:2.4.2}
  \begin{align}
    {\bar J_{{\rm{tp}}}} &= \frac{1}{{{k_f} - {k_0} + 1}}\sum\limits_{k = {k_0}}^{{k_f}} {{J_{{\rm{tp}}}}(k)} \label{eq:2.4.2a}\\
    {\bar J_{{\rm{ep}}}} &= \frac{1}{{{k_f} - {k_0} + 1}}\sum\limits_{k = {k_0}}^{{k_f}} {{J_{{\rm{ep}}}}(k)} \label{eq:2.4.2b}
  \end{align}
\end{subequations}
where ${J_{{\rm{tp}}}}(k)$ and ${J_{{\rm{ep}}}}(k)$ defined as follows denoting the instantaneous tracking and economic indices at time $k$:
\[\begin{array}{l}
    {J_{{\rm{tp}}}}(k) = {\left\| {{\boldsymbol{x}}(k) - {{\boldsymbol{x}}_{\rm{s}}}} \right\|_{\boldsymbol{Q}}^2 + \left\| {{\boldsymbol{u}}(k) - {{\boldsymbol{u}}_{\rm{s}}}} \right\|_{\boldsymbol{R}}^2} \\
    {J_{{\rm{ep}}}}(k) = {E({\boldsymbol{x}}(k),{\boldsymbol{u}}(k))}
\end{array}\]

\section{Motivating example}\label{sec:3}

\begin{figure}
    \centering
    \includegraphics[width=0.5\textwidth,angle=0]{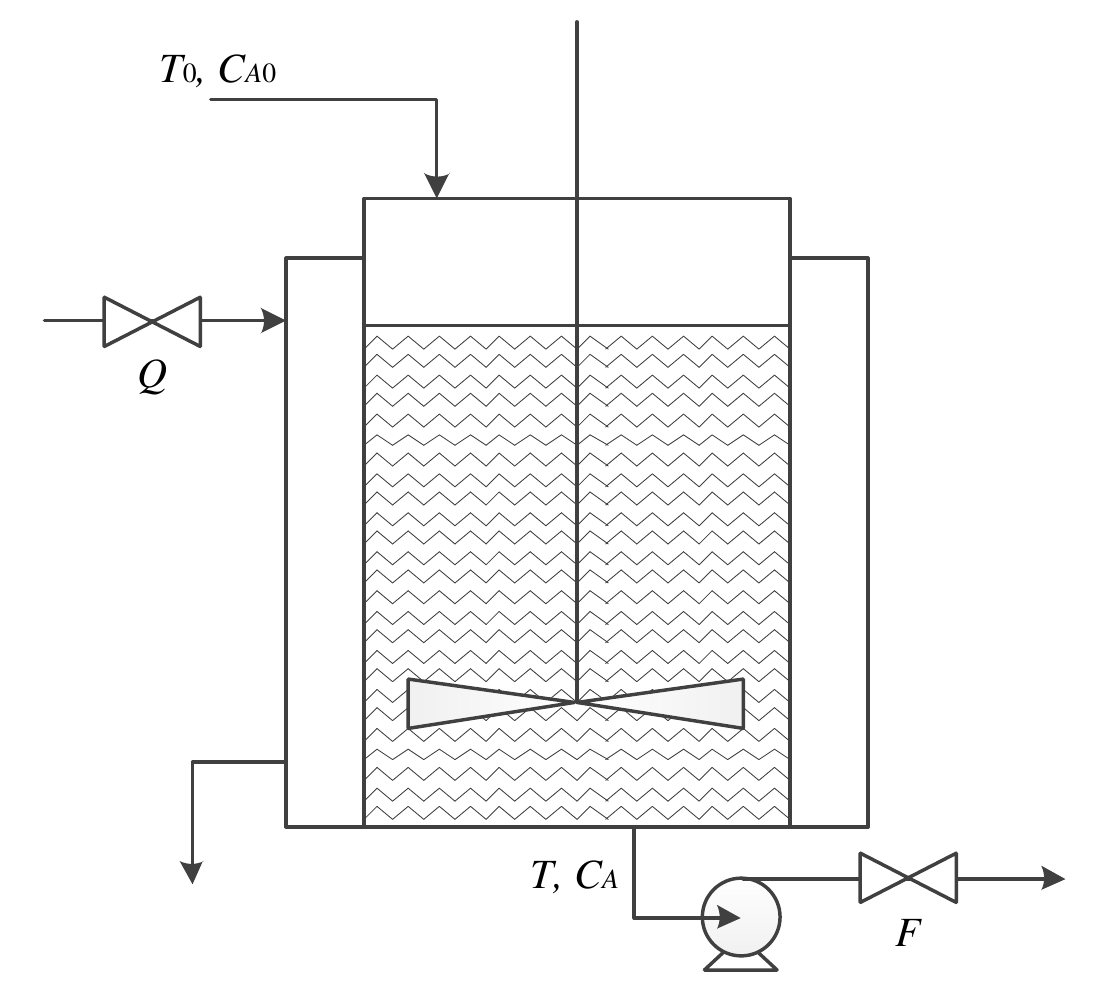}
    \caption{Schematic diagram of the CSTR}\label{fig:3.0.1}
\end{figure}

We consider a nonlinear non-isothermal continuous stirred tank reactor (CSTR) as shown in Figure \ref{fig:3.0.1} to illustrate the motivations of this work. The dynamics of the CSTR are described as follows \cite{henson1997nonlinear}:
\begin{subequations}\label{eq:3.0.1}
  \begin{align}
    &\frac{{d{C_A}}}{{dt}} = \frac{F}{{{V_R}}}({C_{A0}} - {C_A}) - {k_0}{e^{ - \frac{E}{{RT}}}}C_A^2 \label{eq:3.0.1a}\\
    &\frac{{dT}}{{dt}} = \frac{F}{{{V_R}}}({T_0} - T) - \frac{{\Delta H{k_0}}}{{{\rho _R}{c_p}}}{e^{ - \frac{E}{{RT}}}}C_A^2 + \frac{Q}{{{\rho _R}{c_p}{V_R}}} \label{eq:3.0.1b}
  \end{align}
\end{subequations}
where ${C_A}$ is the molar concentration of the reactant $A$; $T$ is the temperature of reactor contents; $F$ is the outlet flow rate of the reactor; $Q$ is the heat rate. The remaining notation definitions and process parameter values are given in Table \ref{tab:3.0.1}. The continuous model is discretized using forward finite difference with a sampling time $\Delta T = 10$ sec. At each sampling time, ${C_A}$ and $T$ are both measured. For this process, the input vector ${\boldsymbol u}=[F~Q]^{\rm T}$, the state vector ${\boldsymbol x}=[C_A ~T]^{\rm T}$, and the output vector ${\boldsymbol y}=[C_A ~T]^{\rm T}$.

\begin{table}
    \caption{Process parameters of the CSTR}\label{tab:3.0.1}
    \centering
    \begin{tabular}{lll}
      \hline
        Symbol & Description &      Value \\
      \hline
           $C_{A0}$ & Feed concentration of A &3.5 ${\rm {kmol/m^3}}$ \\

            $T_0$ & Feedstock temperature &      300 $\rm K$ \\

            $V_R$ & Reactor fluid volume &     1.0 $\rm{m^3}$ \\

             $E$ & Activation energy & 5.0e4 $\rm{kJ/kmol}$ \\

            $k_0$ & Pre-exponential rate factor & 8.46e6 $\rm{m^3/kmol \cdot h}$ \\

            $\Delta H$ & Reaction enthalpy change & -1.16e4 $\rm{kJ/kmol}$ \\

            $c_p$ & Heat capacity & 0.231 $\rm{kJ/kg \cdot K}$ \\

           $\rho _R$ &    Density & 1000 $\rm{kg/m^3}$ \\

             $R$ & Gas constant & 8.314 $\rm{kJ/kmol \cdot K}$ \\
      \hline
    \end{tabular}
\end{table}

In this section, we configure two group experiments to show the different control performance of EMPC and tracking MPC in the presence of noise. In each group, EMPC and tracking MPC finally arrive at the same optimal steady state. In group one, the tracking and economic indices expressed in Eqs. (\ref{eq:2.2.1}), (\ref{eq:2.3.1}), and (\ref{eq:2.4.2}) are set as:
\begin{subequations}\label{eq:3.0.2}
\begin{align}
  {J_{\rm{tp}}} &= {\left\| {{\boldsymbol{x}}(k) - {{\boldsymbol{x}}_{\rm{s}}}} \right\|_{\boldsymbol{Q}}^2 + \left\| {{\boldsymbol{u}}(k) - {{\boldsymbol{u}}_{\rm{s}}}} \right\|_{\boldsymbol{R}}^2} \label{eq:3.0.2a}\\
  {J_{\rm{ep}}} &= {{x_1}{u_1} + 10^5{x_2} + \frac{2}{{{{({x_1} - 1)}^2}}} + \frac{10^4}{{{{({x_2} - 600)}^2}}}} \label{eq:3.0.2b}
\end{align}
\end{subequations}
where the last two terms in Eq. (\ref{eq:3.0.2b}) are known as barrier functions to push the corresponding variables away from their boundaries, $\boldsymbol Q$ and $\boldsymbol R$ are configured as follows to avoid the influence of parameter tuning in tracking MPC.
\[\begin{array}{l}
    {\boldsymbol{Q}} = {\rm{diag}}\left\{ {1/x_{1,{\rm{s}}}^2,1/x_{2,{\rm{s}}}^2, \cdots ,1/x_{n,{\rm{s}}}^2} \right\}\\
    {\boldsymbol{R}} = {\rm{diag}}\left\{ {1/u_{1,{\rm{s}}}^2,1/u_{2,{\rm{s}}}^2, \cdots ,1/u_{m,{\rm{s}}}^2} \right\}
\end{array}\]

The constraints considered in the experiments are shown as follows:
\begin{equation}\label{eq:3.0.3}
  1 \le {x_1} \le 3, ~600 \le {x_2} \le 1000
\end{equation}

The process and measurement noise added on states and measurements is as follows:
\begin{equation}\label{eq:3.0.4}
  \begin{array}{l}
    {w_1}\sim {\rm N}(0,0.00001),{v_1}\sim {\rm N}(0,0.01)\\
    {w_2}\sim {\rm N}(0,0.00009),{v_2}\sim {\rm N}(0,0.09)
  \end{array}
\end{equation}
where ${\rm N}(\mu,\sigma^2)$ denotes the normal distribution function with mean $\mu$ and variance $\sigma^2$.

Group two adopts the same constraint and noise configuration except for a different economic index shown as below:
\begin{equation}\label{eq:3.0.5}
    \begin{array}{l}
      {J_{\rm{ep}}} = 10^2 x_1 + 10^{-4} x_2 u_1 - 2 \times 10^2 \ln(\frac{x_2-600}{200}) - 10^{-3} \ln(\frac{1000-x_2}{200})\\
      \kern 125pt - 10^{-3} \ln(3-x_1) - 1.5 \times 10^3 \ln(x_1-1)
    \end{array}
\end{equation}

Table \ref{tab:3.1.1} shows the average tracking and economic performance of tracking MPC and EMPC. It is shown that sometimes EMPC achieves better economic performance than tracking MPC, sometimes not. Therefore, an unified conclusion about which one is better cannot be directly given.

\begin{table}[H]
    \caption{Average performance indices of EMPC and tracking MPC in the presence of noise}\label{tab:3.1.1}
    \centering
    \begin{tabular}{cllll}
      \hline
          Group & Controller     & Steady state ${\boldsymbol x}_{\rm s}$ & $\bar J_{\rm {ep}}$ & $\bar J_{\rm {tp}}$ \\
      \hline
      \multirow{2}{*}{1} &     Economic MPC & $[1.1601~615.7373]^{\rm T}$ & 453.5738 & 2.0905e-03 \\

          & Tracking MPC & $[1.1601~615.7373]^{\rm T}$ & 448.9241 & 1.0275e-03 \\
      \multirow{2}{*}{2} &     Economic MPC & $[2.7284~631.7380]^{\rm T}$ & 199.8444 & 7.1578e-03 \\

          & Tracking MPC & $[2.7284~631.7380]^{\rm T}$ & 202.5862 & 6.7702e-03 \\
      \hline
    \end{tabular}
\end{table}

Therefore, in this work we aim at providing an approach to analyze the actually accessible performance and giving guidance on which controller should be employed for a given chemical process. We start from the Gaussian noise and propose a sensitivity based dynamic performance assessment approach to pre-evaluate the actual performance of tracking MPC and EMPC. The results are also helpful to conduct boundary moving (i.e. back off) and target moving when $\boldsymbol x$ is located on or close to its boundaries, to guarantee the safety of controlled systems.

\section{Sensitivity-based dynamic performance assessment}\label{sec:4}
\begin{figure}
    \centering
    \includegraphics[width=0.55\textwidth,angle=0]{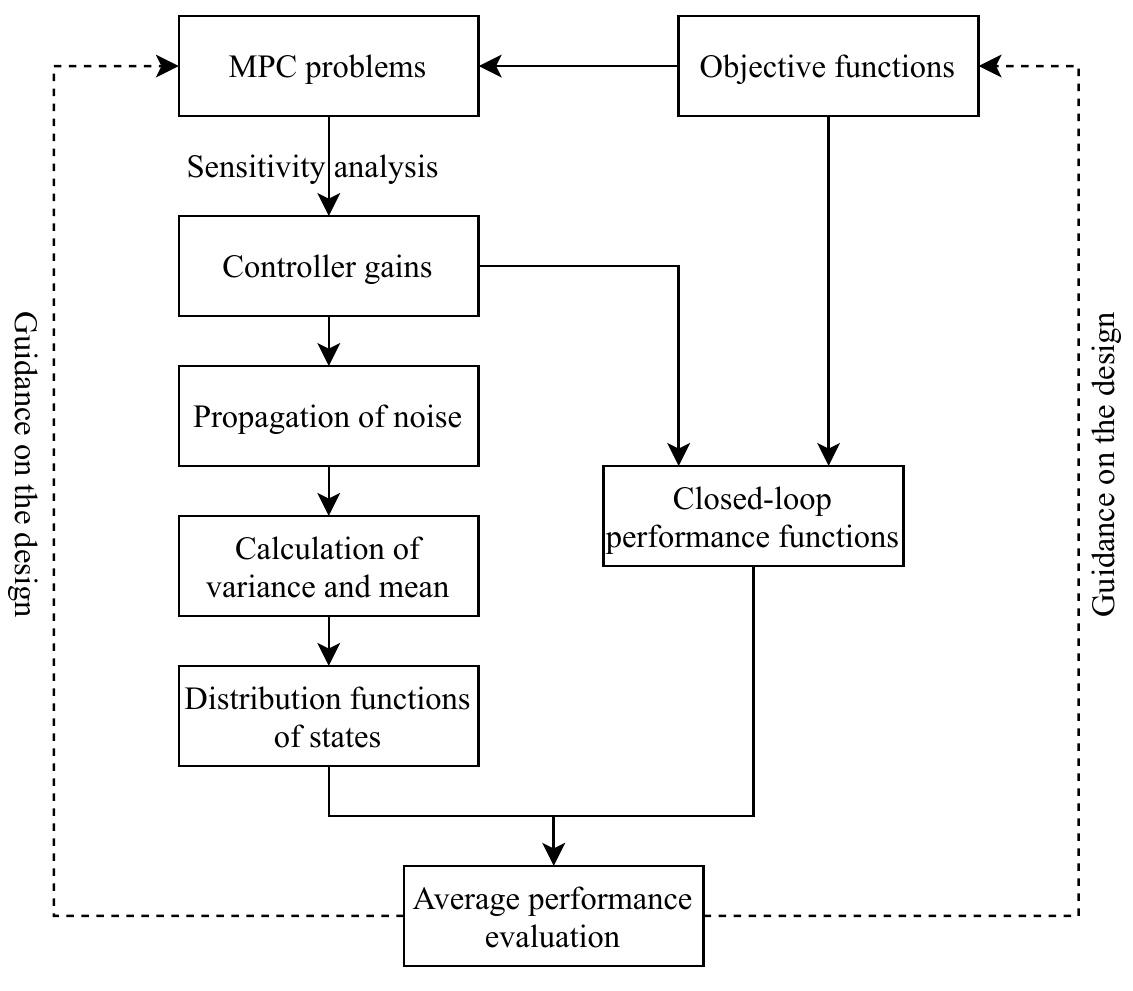}
    \caption{Implementation procedure of the proposed performance assessment approach}\label{fig:4.0.1}
\end{figure}

Sensitivity analysis is to evaluate the sensitivity of the optimal solution with respect to given parameters in constrained optimization problems \cite{evers1980sensitivity,wachter2006implementation,huang2009advanced,andersson2018sensitivity}. For a MPC problem, we usually convert it to a standard nonlinear programming formulation, then analyze its sensitivity information\cite{wachter2006implementation,andersson2018sensitivity}. This technology has been applied in industrial process control to reduce the computational burden as well as the computational time \cite{pirnay2012optimal,lopez2012moving,biegler2013survey}. In this work, we use it to analyze the sensitivity of the input $\boldsymbol u$ with respect to the state ${\boldsymbol x}$ and use it as the gain of the MPC controller. Using the calculated controller gain, we can derive the propagation of noise in the closed-loop system and get a noise transfer model. Based on the model, we can calculate the mean and variance of each state which reflect the distribution information of state variables. Finally, we can evaluate the economic and tracking performance of the controlled system by integrating the closed-loop performance functions which include only states and the calculated distribution functions of states. Figure \ref{fig:4.0.1} shows the implementation procedure of the proposed performance assessment approach. The algorithm proposed in \cite{andersson2018sensitivity} is employed to carry out sensitivity analysis using CasADi with Python in the work \cite{andersson2019casadi}.

\subsection{Sensitivity based controller gain}\label{sec:4.1}
First, we calculate the controller gain of a MPC controller using sensitivity analysis in the section. We treat the initial state ${\boldsymbol x}(0)$ as a sensitivity parameter. Future state ${\boldsymbol x}(1:N)$ and input ${\boldsymbol u}(0:N-1)$ are decision variables.

Let us reform the MPC problems (\ref{eq:2.2.1}) and (\ref{eq:2.3.1}) to the following standard nonlinear programming (NLP) formulation:
\begin{equation}\label{eq:4.1.1}
  \begin{array}{l}
    \mathop {\min }\limits_{{\boldsymbol{d}},{\boldsymbol{p}}} f({\boldsymbol{d}},{\boldsymbol{p}})\\
    {\rm{s}}{\rm{.t}}{\rm{.}}\\
    {\boldsymbol{p}} = {{\boldsymbol{p}}_0},{{\boldsymbol{d}}^{{\rm{lb}}}} \le {\boldsymbol{d}} \le {{\boldsymbol{d}}^{{\rm{ub}}}},{{\boldsymbol{g}}^{{\rm{lb}}}} \le g({\boldsymbol{d}},{\boldsymbol{p}}) \le {{\boldsymbol{g}}^{{\rm{ub}}}}
\end{array}
\end{equation}
where
\[\begin{array}{l}
  {\boldsymbol{d}}: = \left[ {{\boldsymbol{x}}(1)^{\rm T} ~{\boldsymbol{x}}(2)^{\rm T} ~\cdots ~{\boldsymbol{x}}(N)^{\rm T} ~{\boldsymbol{u}}(0)^{\rm T} ~{\boldsymbol{u}}(1)^{\rm T} ~\cdots ~{\boldsymbol{u}}(N - 1)^{\rm T}} \right]^{\rm T}\\
  {\boldsymbol{p}}: = \left[ {{\boldsymbol{x}}(0)^{\rm T} ~{\boldsymbol{\tilde p}}^{\rm T}} \right]^{\rm T}
\end{array}\]
and $N$ is the prediction horizon; ${\boldsymbol{x}}(k)$ and ${\boldsymbol{u}}(k)$ denote the state and input vectors at time $k$, respectively; ${\boldsymbol d}$ denotes the decision variable, which is formed by the state vector $\boldsymbol x(\cdot)$ and the input vector $\boldsymbol u(\cdot)$; ${\boldsymbol p}$ includes the parameters which need to conduct sensitivity analysis and in our work it only contains the initial state ${\boldsymbol{x}}{(0)}$, note that if necessary other sensitivity parameter ${\boldsymbol{\tilde{p}}}$ can also be included; ${\boldsymbol p}_0$ denotes the nominal value of $\boldsymbol p$; $f(\cdot)$ and $g(\cdot)$ denote the objective and constraint functions, respectively; ${\boldsymbol d}^{\rm{lb}}$, ${\boldsymbol d}^{\rm{ub}}$, ${\boldsymbol g}^{\rm{lb}}$, and ${\boldsymbol g}^{\rm{ub}}$ denote the lower and upper boundaries of $\boldsymbol{d}$ and $g(\cdot)$, respectively.

To solve the above constrained optimization problem, we need to introduce Lagrangian multipliers ${\boldsymbol{\lambda }}_{\boldsymbol{d}}^{{\rm{lb}}}$, ${\boldsymbol{\lambda }}_{\boldsymbol{d}}^{{\rm{ub}}}$, ${\boldsymbol{\lambda }}_g^{{\rm{lb}}}$, ${\boldsymbol{\lambda }}_g^{{\rm{ub}}}$, and ${{\boldsymbol{\lambda }}_{\boldsymbol{p}}}$ to convert it to an unconstrained optimization problem as follows:
\begin{equation}\label{eq:4.1.2}
  \begin{array}{l}
    \tilde L({\boldsymbol{d}},{\boldsymbol{p}},{\boldsymbol{\lambda }}_{\boldsymbol{d}}^{{\rm{lb}}},{\boldsymbol{\lambda }}_{\boldsymbol{d}}^{{\rm{ub}}},{\boldsymbol{\lambda }}_g^{{\rm{lb}}},{\boldsymbol{\lambda }}_g^{{\rm{ub}}},{{\boldsymbol{\lambda }}_{\boldsymbol{p}}}) = {\mathop{f}\nolimits} ({\boldsymbol{d}},{\boldsymbol{p}}) + {{\boldsymbol{\lambda }}_{\boldsymbol{p}}}({\boldsymbol{p}} - {{\boldsymbol{p}}_0}) - {\boldsymbol{\lambda }}_{\boldsymbol{d}}^{{\rm{lb}}}({\boldsymbol{d}} - {{\boldsymbol{d}}^{{\rm{lb}}}}) - {\boldsymbol{\lambda }}_{\boldsymbol{d}}^{{\rm{ub}}}({{\boldsymbol{d}}^{{\rm{ub}}}} - {\boldsymbol{d}})\\
    \kern 194pt  - {\boldsymbol{\lambda }}_g^{{\rm{lb}}}(g({\boldsymbol{d}},{\boldsymbol{p}}) - {{\boldsymbol{g}}^{{\rm{lb}}}}) - {\boldsymbol{\lambda }}_g^{{\rm{ub}}}({{\boldsymbol{g}}^{{\rm{ub}}}} - g({\boldsymbol{d}},{\boldsymbol{p}}))
  \end{array}
\end{equation}

The Karush-Kuhn-Tucker(KKT) conditions for the above unconstrained problem (\ref{eq:4.1.2}) is as follows:
\begin{equation}\label{eq:4.1.3}
  \begin{array}{l}
    {\nabla _{\boldsymbol{d}}}\tilde L({\boldsymbol{d}},{\boldsymbol{p}},{\boldsymbol{\lambda }}_{\boldsymbol{d}}^{{\rm{lb}}},{\boldsymbol{\lambda }}_{\boldsymbol{d}}^{{\rm{ub}}},{\boldsymbol{\lambda }}_g^{{\rm{lb}}},{\boldsymbol{\lambda }}_g^{{\rm{ub}}},{{\boldsymbol{\lambda }}_{\boldsymbol{p}}}) = {\boldsymbol{0}}\\
    {\nabla _{\boldsymbol{p}}}\tilde L({\boldsymbol{d}},{\boldsymbol{p}},{\boldsymbol{\lambda }}_{\boldsymbol{d}}^{{\rm{lb}}},{\boldsymbol{\lambda }}_{\boldsymbol{d}}^{{\rm{ub}}},{\boldsymbol{\lambda }}_g^{{\rm{lb}}},{\boldsymbol{\lambda }}_g^{{\rm{ub}}},{{\boldsymbol{\lambda }}_{\boldsymbol{p}}}) = {\boldsymbol{0}}\\
    {\boldsymbol{p}} = {{\boldsymbol{p}}_0},{\kern 1pt} {\kern 1pt} {\kern 1pt} {\kern 1pt} {\kern 1pt} {\boldsymbol{d}} - {{\boldsymbol{d}}^{{\rm{lb}}}} \ge {\boldsymbol{0}},{\kern 1pt} {\kern 1pt} {\kern 1pt} {\kern 1pt} {\kern 1pt} {{\boldsymbol{d}}^{{\rm{ub}}}} - {\boldsymbol{d}} \ge {\boldsymbol{0}}\\
    g({\boldsymbol{d}},{\boldsymbol{p}}) - {{\boldsymbol{g}}^{{\rm{lb}}}} \ge {\boldsymbol{0}},{\kern 1pt} {\kern 1pt} {\kern 1pt} {\kern 1pt} {{\boldsymbol{g}}^{{\rm{ub}}}} - g({\boldsymbol{d}},{\boldsymbol{p}}) \ge {\boldsymbol{0}}\\
    {\boldsymbol{\lambda }}_{\boldsymbol{d}}^{{\rm{lb}}} \ge {\boldsymbol{0}},{\kern 1pt} {\kern 1pt} {\kern 1pt} {\kern 1pt} {\kern 1pt} {\boldsymbol{\lambda }}_{\boldsymbol{d}}^{{\rm{ub}}} \ge {\boldsymbol{0}},{\kern 1pt} {\kern 1pt} {\kern 1pt} {\kern 1pt} {\kern 1pt} {\boldsymbol{\lambda }}_g^{{\rm{lb}}} \ge {\boldsymbol{0}},{\kern 1pt} {\kern 1pt} {\kern 1pt} {\kern 1pt} {\kern 1pt} {\boldsymbol{\lambda }}_g^{{\rm{ub}}} \ge {\boldsymbol{0}}\\
    {\boldsymbol{\lambda }}_{\boldsymbol{d}}^{{\rm{lb}}}({\boldsymbol{d}} - {{\boldsymbol{d}}^{{\rm{lb}}}}) = {\boldsymbol{0}},{\kern 1pt} {\kern 1pt} {\kern 1pt} {\kern 1pt} {\kern 1pt} {\boldsymbol{\lambda }}_{\boldsymbol{d}}^{{\rm{ub}}}({{\boldsymbol{d}}^{{\rm{ub}}}} - {\boldsymbol{d}}) = {\boldsymbol{0}}\\
    {\boldsymbol{\lambda }}_g^{{\rm{lb}}}(g({\boldsymbol{d}},{\boldsymbol{p}}) - {{\boldsymbol{g}}^{{\rm{lb}}}}) = {\boldsymbol{0}},{\kern 1pt} {\kern 1pt} {\kern 1pt} {\kern 1pt} {\kern 1pt} {\boldsymbol{\lambda }}_g^{{\rm{ub}}}({{\boldsymbol{g}}^{{\rm{ub}}}} - g({\boldsymbol{d}},{\boldsymbol{p}})) = {\boldsymbol{0}}
  \end{array}
\end{equation}

To simplify the above KKT conditions (\ref{eq:4.1.3}), we introduce ${{\boldsymbol{\lambda }}_{\boldsymbol{d}}} = {\boldsymbol{\lambda }}_{\boldsymbol{d}}^{{\rm{ub}}} - {\boldsymbol{\lambda }}_{\boldsymbol{d}}^{{\rm{lb}}}$ and ${{\boldsymbol{\lambda }}_g} = {\boldsymbol{\lambda }}_g^{{\rm{ub}}} - {\boldsymbol{\lambda }}_g^{{\rm{lb}}}$ and a new Lagrangian function with all linear terms eliminated can be obtained as follows:
\begin{equation}\label{eq:4.1.4}
  L({\boldsymbol{d}},{\boldsymbol{p}},{{\boldsymbol{\lambda }}_g}) = f({\boldsymbol{d}},{\boldsymbol{p}}) + {{\boldsymbol{\lambda }}_g}g({\boldsymbol{d}},{\boldsymbol{p}})
\end{equation}

The new equivalent KKT conditions can be formulated as below:
\begin{subequations}\label{eq:4.1.5}
  \begin{align}
    &{\nabla _{\boldsymbol{d}}}L({\boldsymbol{d}},{\boldsymbol{p}},{{\boldsymbol{\lambda }}_g}) + {{\boldsymbol{\lambda }}_{\boldsymbol{d}}} = {\boldsymbol{0}} \label{eq:4.1.5a}\\
    &{\nabla _{\boldsymbol{p}}}L({\boldsymbol{d}},{\boldsymbol{p}},{{\boldsymbol{\lambda }}_g}) + {{\boldsymbol{\lambda }}_{\boldsymbol{p}}} = {\boldsymbol{0}} \label{eq:4.1.5b}\\
    &{{\boldsymbol{d}}^{{\rm{lb}}}} \le {\boldsymbol{d}} \le {{\boldsymbol{d}}^{{\rm{ub}}}},{{\boldsymbol{g}}^{{\rm{lb}}}} \le g({\boldsymbol{d}},{\boldsymbol{p}}) \le {{\boldsymbol{g}}^{{\rm{ub}}}} \label{eq:4.1.5c}\\
    &{\boldsymbol{I}}_{\boldsymbol{d}}^0{{\boldsymbol{\lambda }}_{\boldsymbol{d}}} + {\boldsymbol{I}}_{\boldsymbol{d}}^{{\rm{ub}}}{{\boldsymbol{d}}^{{\rm{ub}}}} + {\boldsymbol{I}}_{\boldsymbol{d}}^{{\rm{lb}}}{{\boldsymbol{d}}^{{\rm{lb}}}} - {{\boldsymbol{I}}_{\boldsymbol{d}}}{\boldsymbol{d}} = {\boldsymbol{0}} \label{eq:4.1.5d}\\
    &{\boldsymbol{I}}_g^0{{\boldsymbol{\lambda }}_g} + {\boldsymbol{I}}_g^{{\rm{ub}}}{{\boldsymbol{g}}^{{\rm{ub}}}} + {\boldsymbol{I}}_g^{{\rm{lb}}}{{\boldsymbol{g}}^{{\rm{lb}}}} - {{\boldsymbol{I}}_g}g({\boldsymbol{d}},{\boldsymbol{p}}) = {\boldsymbol{0}} \label{eq:4.1.5e}
  \end{align}
\end{subequations}
where ${\boldsymbol{I}}_ \bullet ^{{\rm{lb}}}$, ${\boldsymbol{I}}_ \bullet ^{{\rm{ub}}}$, ${{\boldsymbol{I}}_ \bullet }$, and ${\boldsymbol{I}}_ \bullet ^0$ are the active constraints flags that denote whether the corresponding constraints are active or not \cite{andersson2018sensitivity}.

Using (\ref{eq:4.1.5a}), we can eliminate ${{\boldsymbol{\lambda }}_{\boldsymbol{d}}}$ from (\ref{eq:4.1.5d}) and get an implicit function $F({\boldsymbol{z}},{\boldsymbol{q}})$ as follows:
\begin{equation}\label{eq:4.1.6}
  F({\boldsymbol{z}},{\boldsymbol{q}}) := \left[ {\begin{array}{*{20}{c}}
  {{\boldsymbol{I}}_{\boldsymbol{d}}^0{\nabla _{\boldsymbol{d}}}L({\boldsymbol{d}},{\boldsymbol{p}},{{\boldsymbol{\lambda }}_g}) + {{\boldsymbol{I}}_{\boldsymbol{d}}}{\boldsymbol{d}} - {\boldsymbol{I}}_{\boldsymbol{d}}^{{\rm{ub}}}{{\boldsymbol{d}}^{{\rm{ub}}}} - {\boldsymbol{I}}_{\boldsymbol{d}}^{{\rm{lb}}}{{\boldsymbol{d}}^{{\rm{lb}}}}}\\
  {{{\boldsymbol{I}}_g}g({\boldsymbol{d}},{\boldsymbol{p}}) - {\boldsymbol{I}}_g^0{{\boldsymbol{\lambda }}_g} - {\boldsymbol{I}}_g^{{\rm{ub}}}{{\boldsymbol{g}}^{{\rm{ub}}}} - {\boldsymbol{I}}_g^{{\rm{lb}}}{{\boldsymbol{g}}^{{\rm{lb}}}}}
  \end{array}} \right] = {\boldsymbol{0}}
\end{equation}
where
\[\begin{array}{l}
  {\boldsymbol{z}}: = [{\boldsymbol{d}}^{\rm T} ~{{\boldsymbol{\lambda }}_g}^{\rm T}]^{\rm T}\\
  {\boldsymbol{q}}: = [{{\boldsymbol{d}}^{{\rm{lb}}}}^{\rm T} ~{{\boldsymbol{d}}^{{\rm{ub}}}}^{\rm T} ~{{\boldsymbol{g}}^{{\rm{lb}}}}^{\rm T} ~{{\boldsymbol{g}}^{{\rm{ub}}}}^{\rm T} ~{\boldsymbol{p}}^{\rm T}]^{\rm T}
\end{array}\]

If we define a corresponding explicit function ${\boldsymbol{z}}: = G({\boldsymbol{q}})$, we can get the sensitivity of $\boldsymbol z$ with respect to $\boldsymbol q$, $\frac{d \boldsymbol z}{d \boldsymbol q}$, using the implicit function theorem \cite{krantz2012implicit}:
\begin{equation}\label{eq:4.1.7}
  \frac{{d \boldsymbol z}}{{d {\boldsymbol{q}}}} =  - {\left[ {\frac{{\partial F}}{{\partial {\boldsymbol{z}}}}} \right]^{ - 1}}\frac{{\partial F}}{{\partial {\boldsymbol{q}}}}
\end{equation}
where
\[\frac{{\partial F}}{{\partial {\boldsymbol{z}}}} = \left[ {\begin{array}{*{20}{c}}
{{\boldsymbol{I}}_{\boldsymbol{d}}^0\nabla _{\boldsymbol{d}}^2L({\boldsymbol{d}},{\boldsymbol{p}},{{\boldsymbol{\lambda }}_g}) + {{\boldsymbol{I}}_{\boldsymbol{d}}}}&{{\boldsymbol{I}}_{\boldsymbol{d}}^0{{\left[ {\frac{{\partial g}}{{\partial {\boldsymbol{d}}}}({\boldsymbol{d}},{\boldsymbol{p}})} \right]}^{\rm{T}}}}\\
{{{\boldsymbol{I}}_g}\frac{{\partial g}}{{\partial {\boldsymbol{d}}}}({\boldsymbol{d}},{\boldsymbol{p}})}&{ - {\boldsymbol{I}}_g^0}
\end{array}} \right]\]
\[\frac{{\partial F}}{{\partial {\boldsymbol{q}}}} = \left[ {\begin{array}{*{20}{c}}
{ - {\boldsymbol{I}}_{\boldsymbol{d}}^{{\rm{lb}}}}&{ - {\boldsymbol{I}}_{\boldsymbol{d}}^{{\rm{ub}}}}&{\boldsymbol{0}}&{\boldsymbol{0}}&{{\boldsymbol{I}}_{\boldsymbol{d}}^0\frac{\partial }{{\partial {\boldsymbol{p}}}}{\nabla _{\boldsymbol{d}}}L({\boldsymbol{d}},{\boldsymbol{p}},{{\boldsymbol{\lambda }}_g})}\\
{\boldsymbol{0}}&{\boldsymbol{0}}&{ - {\boldsymbol{I}}_g^{{\rm{lb}}}}&{ - {\boldsymbol{I}}_g^{{\rm{ub}}}}&{{{\boldsymbol{I}}_g}\frac{\partial }{{\partial {\boldsymbol{p}}}}g({\boldsymbol{d}},{\boldsymbol{p}})}
\end{array}} \right]\]

In MPC, the controller gain ${{\boldsymbol{K}}^{{\rm{MPC}}}}$ can be regarded as the sensitivity of ${\boldsymbol u}(0)$ with respect to ${\boldsymbol x}(0)$, which is a sub-block of the above calculated sensitivity matrix $\frac{d \boldsymbol z}{d \boldsymbol q}$. Therefore, we can get the following controller gain ${{\boldsymbol{K}}^{{\rm{MPC}}}}$ from the calculated $\frac{d \boldsymbol z}{d \boldsymbol q}$.
\begin{equation}\label{eq:4.1.8}
  {{\boldsymbol{K}}^{{\rm{MPC}}}} = {\left.\frac{{\partial {\boldsymbol{u}}(0)}}{{\partial {\boldsymbol{x}}(0)}}\right|_{{{\boldsymbol{x}}_0}}}
\end{equation}
where ${{\boldsymbol{K}}^{{\rm{MPC}}}}$ is the controller gain of MPC and ${\boldsymbol{x}}_0$ denotes the nominal value of ${\boldsymbol{x}}(0)$.

Finally, the optimal feedback control law of MPC can be formulated as below:
\begin{equation}\label{eq:4.1.9}
  {\boldsymbol{u}}(0) = {{\boldsymbol{u}}_{\rm{0}}} + {{\boldsymbol{K}}^{\rm{MPC}}}({\boldsymbol{x}}(0) - {{\boldsymbol{x}}_{\rm{0}}})
\end{equation}
where ${\boldsymbol{u}}_0$ denotes the nominal optimal solution, ${\boldsymbol x}(0)$ is the actual state value at time 0, and ${\boldsymbol u}(0)$ is the actual optimal input at time 0.

\subsection{The propagation of process and measurement noise}\label{sec:4.2}
Based on the above calculated controller gain, we can derive the noise propagation of noise in a closed-loop controlled system. Assume that Gaussian process and measurement noise is added on states and measurements and all states are directly measured, the complete closed-loop system around the given steady state can be described as follows:
\begin{subequations}\label{eq:4.2.1}
  \begin{align}
    &{\boldsymbol{y}}(k) = {{\boldsymbol{x}}}(k) + {\boldsymbol{v}}(k) \label{eq:4.2.1a}\\
    &{{\boldsymbol{x}}_{\rm{m}}}(k) = {\boldsymbol{y}}(k) \label{eq:4.2.1b}\\
    &{\boldsymbol{u}}(k) = {{\boldsymbol{u}}_{\rm{s}}} + {{\boldsymbol{K}}^{{\rm{MPC}}}}({{\boldsymbol{x}}_{\rm{m}}}(k) - {{\boldsymbol{x}}_{\rm{s}}}) \label{eq:4.2.1c}\\
    &{{\boldsymbol{x}}}(k + 1) = F({{\boldsymbol{x}}}(k),{\boldsymbol{u}}(k)) + {\boldsymbol{w}}(k) \label{eq:4.2.1d}
  \end{align}
\end{subequations}
where ${\boldsymbol{x}}(k)$ and ${\boldsymbol{x}}_{\rm m}(k)$ denote the actual and measured state values at time $k$, respectively; ${\boldsymbol x}(k)$ is usually supposed to be unknown and ${\boldsymbol{x}}_{\rm m}(k)$ instead of ${\boldsymbol x}(k)$ is used to compute the optimal input ${\boldsymbol{u}}(k)$.

Finally, Eq. (\ref{eq:4.2.2}) can be achieved by successively substituting (\ref{eq:4.2.1c}), (\ref{eq:4.2.1b}), and (\ref{eq:4.2.1a}) into (\ref{eq:4.2.1d}):
\begin{equation}\label{eq:4.2.2}
  \begin{array}{l}
    {{\boldsymbol{x}}}(k + 1) = F({{\boldsymbol{x}}}(k),{{\boldsymbol{u}}_{\rm{s}}} + {{\boldsymbol{K}}^{{\rm{MPC}}}}({{\boldsymbol{x}}}(k) + {\boldsymbol{v}}(k) - {{\boldsymbol{x}}_{\rm{s}}})) + {\boldsymbol{w}}(k)
  \end{array}
\end{equation}
It is clear that Eq. (\ref{eq:4.2.2}) only contains the actual state ${\boldsymbol{x}}(\cdot)$, process noise ${\boldsymbol{w}}(k)$, and measurement noise ${\boldsymbol{v}}(k)$, which reveals the propagation of process and measurement noise in a closed-loop system.

\subsection{Calculation of variance and mean}\label{sec:4.3}
Based on the above noise transfer function, we can calculate the mean and variance of each variable, which can be used to get the approximate distribution of states in a well-controlled system. However, although the propagation of process and measurement noise in a closed-loop controlled system has been revealed as Eq. (\ref{eq:4.2.2}), calculating the variances and means of relevant variables is still rather challenging. To simplify the calculation, Taylor expansion is introduced to approximate the nonlinear model (\ref{eq:4.2.2}) around the given steady state $({\boldsymbol{u}}_{\rm s},{\boldsymbol{x}}_{\rm s})$.

Let us denote Eq. (\ref{eq:4.2.2}) by follows:
\begin{equation}\label{eq:4.3.1}
  {{\boldsymbol{x}}}(k + 1): = \varphi ({{\boldsymbol{x}}}(k),{\boldsymbol{w}}(k),{\boldsymbol{v}}(k))
\end{equation}

The first-order Taylor expansion of (\ref{eq:4.2.2}) will be as follows:
\begin{equation}\label{eq:4.3.2}
  {{\boldsymbol{x}}}(k + 1) = \varphi ({{\boldsymbol{x}}_{\rm{s}}},{{\boldsymbol{w}}_{\rm{s}}},{{\boldsymbol{v}}_{\rm{s}}}) + {\left. \frac{{\partial \varphi }}{{\partial {{\boldsymbol{x}}}}}\right|_{{{\boldsymbol{x}}_{\rm{s}}},{{\boldsymbol{w}}_{\rm{s}}},{{\boldsymbol{v}}_{\rm{s}}}}}({{\boldsymbol{x}}}(k) - {{\boldsymbol{x}}_{\rm{s}}}) + {\left. \frac{{\partial \varphi }}{{\partial {{\boldsymbol{w}}}}}\right|_{{{\boldsymbol{x}}_{\rm{s}}},{{\boldsymbol{w}}_{\rm{s}}},{{\boldsymbol{v}}_{\rm{s}}}}}({\boldsymbol{w}}(k) - {{\boldsymbol{w}}_{\rm{s}}}) + {\left. \frac{{\partial \varphi }}{{\partial {{\boldsymbol{v}}}}}\right|_{{{\boldsymbol{x}}_{\rm{s}}},{{\boldsymbol{w}}_{\rm{s}}},{{\boldsymbol{v}}_{\rm{s}}}}}({\boldsymbol{v}}(k) - {{\boldsymbol{v}}_{\rm{s}}})
\end{equation}

If we define
\begin{equation}\label{eq:4.3.3}
  \begin{array}{l}
    {{\boldsymbol{K}}_{{\boldsymbol{x}}2{\boldsymbol{x}}}}: = {\left. \frac{{\partial \varphi }}{{\partial {{\boldsymbol{x}}}}}\right|_{{{\boldsymbol{x}}_{\rm{s}}},{{\boldsymbol{w}}_{\rm{s}}},{{\boldsymbol{v}}_{\rm{s}}}}}\\
    {{\boldsymbol{K}}_{{\boldsymbol{x}}2{\boldsymbol{w}}}}: = {\left. \frac{{\partial \varphi }}{{\partial {{\boldsymbol{w}}}}}\right|_{{{\boldsymbol{x}}_{\rm{s}}},{{\boldsymbol{w}}_{\rm{s}}},{{\boldsymbol{v}}_{\rm{s}}}}}\\
    {{\boldsymbol{K}}_{{\boldsymbol{x}}2{\boldsymbol{v}}}}: = {\left. \frac{{\partial \varphi }}{{\partial {{\boldsymbol{v}}}}}\right|_{{{\boldsymbol{x}}_{\rm{s}}},{{\boldsymbol{w}}_{\rm{s}}},{{\boldsymbol{v}}_{\rm{s}}}}}
  \end{array}
\end{equation}

Eq. (\ref{eq:4.3.2}) can be expressed as follows:
\begin{equation}\label{eq:4.3.4}
  {{\boldsymbol{x}}}(k + 1) = \varphi ({{\boldsymbol{x}}_{\rm{s}}},{{\boldsymbol{w}}_{\rm{s}}},{{\boldsymbol{v}}_{\rm{s}}}) + {{\boldsymbol{K}}_{{\boldsymbol{x}}2{\boldsymbol{x}}}}({{\boldsymbol{x}}}(k) - {{\boldsymbol{x}}_{\rm{s}}}) + {{\boldsymbol{K}}_{{\boldsymbol{x}}2{\boldsymbol{w}}}}({\boldsymbol{w}}(k) - {{\boldsymbol{w}}_{\rm{s}}}) + {{\boldsymbol{K}}_{{\boldsymbol{x}}2{\boldsymbol{v}}}}({\boldsymbol{v}}(k) - {{\boldsymbol{v}}_{\rm{s}}})
\end{equation}

Based on Eq. (\ref{eq:4.3.4}), we can achieve the following covariance matrix equation:
\begin{equation}\label{eq:4.3.5}
  {\boldsymbol\Sigma} _{{{\boldsymbol{x}}}} = {{\boldsymbol{K}}_{{\boldsymbol{x}}2{\boldsymbol{x}}}} {\boldsymbol\Sigma} _{{{\boldsymbol{x}}}} {\boldsymbol{K}}_{{\boldsymbol{x}}2{\boldsymbol{x}}}^{\rm{T}} + {{\boldsymbol{K}}_{{\boldsymbol{x}}2{\boldsymbol{w}}}}{\boldsymbol\Sigma} _{\boldsymbol{w}}{\boldsymbol{K}}_{{\boldsymbol{x}}2{\boldsymbol{w}}}^{\rm{T}} + {{\boldsymbol{K}}_{{\boldsymbol{x}}2{\boldsymbol{v}}}}{\boldsymbol\Sigma} _{\boldsymbol{v}}{\boldsymbol{K}}_{{\boldsymbol{x}}2{\boldsymbol{v}}}^{\rm{T}}
\end{equation}
where ${\boldsymbol\Sigma} _{{{\boldsymbol{x}}}}$, ${\boldsymbol\Sigma} _{\boldsymbol{w}}$, and ${\boldsymbol\Sigma} _{\boldsymbol{v}}$ denote the covariance matrices of ${\boldsymbol{x}}$, $\boldsymbol w$, and $\boldsymbol v$, respectively; ${\boldsymbol\Sigma} _{\boldsymbol{w}}$ and ${\boldsymbol\Sigma} _{\boldsymbol{v}}$ are supposed to be known.

We can also achieve the following mean vector equation:
\begin{equation}\label{eq:4.3.6}
  {{\boldsymbol{\mu}}_{{{\boldsymbol{x}}}}} = \varphi ({{\boldsymbol{x}}_{\rm{s}}},{{\boldsymbol{w}}_{\rm{s}}},{{\boldsymbol{v}}_{\rm{s}}}) = {{\boldsymbol{x}}_{\rm{s}}}
\end{equation}
where ${\boldsymbol {\mu_x}}$ denote the mean vector of $\boldsymbol x$.

Similarly, the covariance matrices of the measured output $\boldsymbol y$, state $\boldsymbol{x}_{\rm m}$, and input $\boldsymbol u$ can also be obtained based on Eqs. (\ref{eq:4.2.1a}), (\ref{eq:4.2.1b}), and (\ref{eq:4.2.1c}) using the same approach:
\begin{equation}\label{eq:4.3.7}
  \begin{array}{l}
    {\boldsymbol\Sigma} _{\boldsymbol{y}} = {{\boldsymbol{K}}_{{\boldsymbol{y}}2{\boldsymbol{x}}}}{\boldsymbol\Sigma} _{{{\boldsymbol{x}}}}{\boldsymbol{K}}_{{\boldsymbol{y}}2{\boldsymbol{x}}}^{\rm{T}} + {{\boldsymbol{K}}_{{\boldsymbol{y}}2{\boldsymbol{v}}}}{\boldsymbol\Sigma} _{\boldsymbol{v}}{\boldsymbol{K}}_{{\boldsymbol{y}}2{\boldsymbol{v}}}^{\rm{T}} \\
    {\boldsymbol\Sigma} _{{{\boldsymbol{x}}_{\rm{m}}}} = {{\boldsymbol{K}}_{{\boldsymbol{x}}2{\boldsymbol{y}}}}{\boldsymbol\Sigma} _{\boldsymbol{y}}{\boldsymbol{K}}_{{\boldsymbol{x}}2{\boldsymbol{y}}}^{\rm{T}} \\
    {\boldsymbol\Sigma} _{\boldsymbol{u}} = {{\boldsymbol{K}}^{{\rm{MPC}}}}{\boldsymbol\Sigma} _{{{\boldsymbol{x}}_{\rm{m}}}}{\left( {{{\boldsymbol{K}}^{{\rm{MPC}}}}} \right)^{\rm{T}}}
  \end{array}
\end{equation}
where ${{\boldsymbol{K}}_{{\bullet}2{\star}}}$ denotes the corresponding sensitivity matrix of $\bullet$ with respect to $\star$.

\subsection{Dynamic performance assessment}\label{sec:4.4}
In this section, we use the above results to obtain the distribution functions of states and the closed-loop performance functions which contains only states. Then, we integrate the distribution functions and the closed-loop performance functions to evaluate the dynamic economic and tracking performance of the controlled systems.

\subsubsection{Distribution function of states}\label{sec:4.4.1}
Each state can be supposed to satisfy normal distribution and all states satisfy joint normal distribution as (\ref{eq:4.4.1}), if only Gaussian process and measurement noise exists in a well controlled system. Note that the measured ${\boldsymbol x}_{\rm m}$ is be used to evaluate the performance in practical engineering instead of $\boldsymbol x$ due to its unpredictability.
\begin{equation}\label{eq:4.4.1}
  {\boldsymbol{x}}_{\rm m}\sim {\rm N}({\boldsymbol{\mu }},{\boldsymbol{\Sigma }})
\end{equation}
where $\boldsymbol{\mu}$ is the mean vector and $\boldsymbol{\Sigma}$ is the covariance matrix.

Therefore, the joint probability density function of states can be expressed as follows:
\begin{equation}\label{eq:4.4.2}
  p({\boldsymbol{x}}_{\rm m}): = \psi ({\boldsymbol{\mu }},{\boldsymbol{\Sigma }},{\boldsymbol{\rho }})
\end{equation}
where $\boldsymbol{\rho}$ denotes the correlation coefficients between different states and can be calculated based on the covariance matrix $\boldsymbol\Sigma$, $\psi(\cdot)$ is the joint normal probability density function.

\subsubsection{Closed-loop performance function and surface}\label{sec:4.4.2}

If we substitute the controller gain ${\boldsymbol{K}}^{\rm {MPC}}$ into the performance indices $J_{\rm ep}(k)$ and $J_{\rm tp}(k)$ shown in (\ref{eq:2.4.2}), we can get the closed-loop economic and tracking performance functions as follows which eliminate inputs and contain states only.
\begin{subequations}\label{eq:4.4.3}
  \begin{align}
    {J_{{\rm{ep}}}}({\boldsymbol{x}}_{\rm m}) &:= {J_{{\rm{ep}}}}({\boldsymbol{x}}_{\rm m},{{\boldsymbol{u}}_{\rm{s}}} + {{\boldsymbol{K}}^{{\rm{MPC}}}}({\boldsymbol{x}}_{\rm m} - {{\boldsymbol x}_{\rm{s}}})) \label{eq:4.4.3a}\\
    {J_{{\rm{tp}}}}({\boldsymbol{x}}_{\rm m}) &:= {J_{{\rm{tp}}}}({\boldsymbol{x}}_{\rm m},{{\boldsymbol{u}}_{\rm{s}}} + {{\boldsymbol{K}}^{{\rm{MPC}}}}({\boldsymbol{x}}_{\rm m} - {{\boldsymbol x}_{\rm{s}}})) \label{eq:4.4.3b}
  \end{align}
\end{subequations}

\begin{figure}
    \centering
    \subfigure[Economic performance surface]{
      \label{subfig:4.4.1a} 
      \includegraphics[width=0.475\textwidth,angle=0]{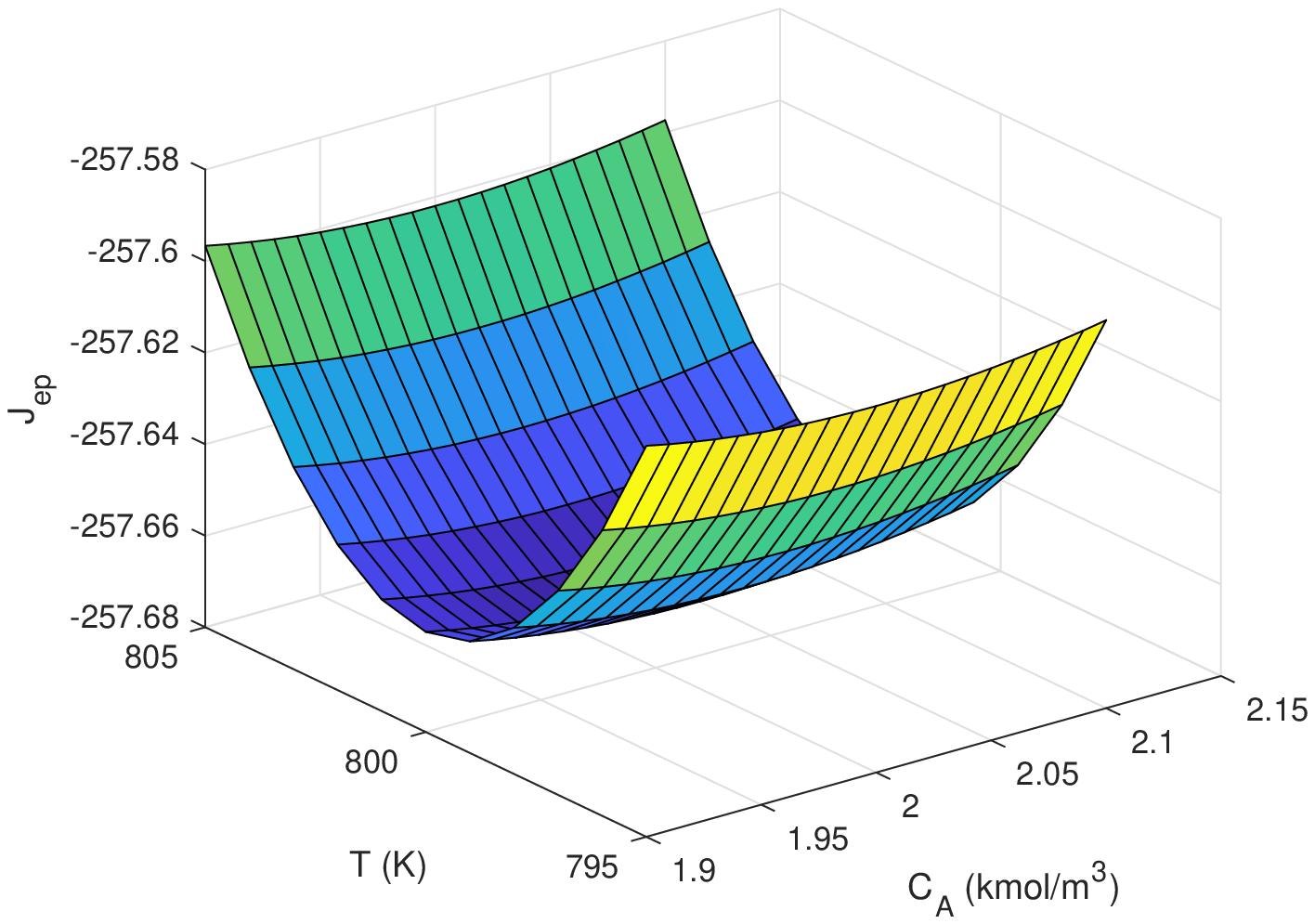}}
    \hspace{0in}
    \subfigure[Tracking performance surface]{
      \label{subfig:4.4.1b} 
      \includegraphics[width=0.475\textwidth,angle=0]{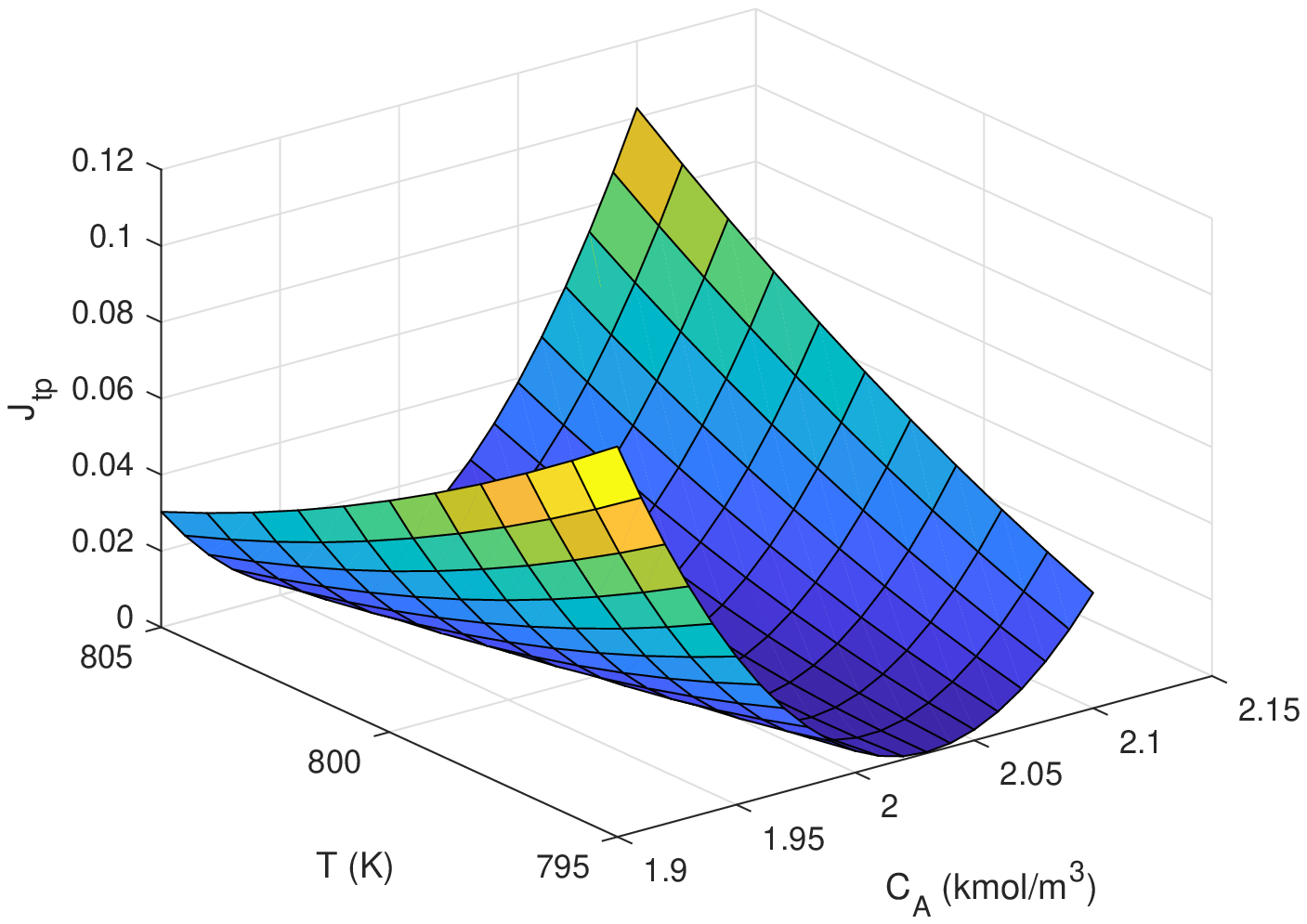}}
    \caption{Example of dynamic economic and tracking performance surfaces}\label{fig:4.4.1}
\end{figure}

The above closed-loop performance functions can be used to evaluate the tracking and economic performance of controlled systems. For better understand their influence, we can plot the dynamic tracking and economic performance surfaces around the given steady state using them, which could intuitively show the sensitivity degree of the performance indices with respect to states. The calculated closed-loop performance functions and plotted surfaces are also helpful for designing a better objective function with lower sensitivity with respect to noise. Figure \ref{fig:4.4.1} shows a simple example of the dynamic performance surfaces, where Figure \ref{subfig:4.4.1a} denotes the dynamic economic performance surface and Figure \ref{subfig:4.4.1b} is the dynamic tracking performance surface.

\subsubsection{Average performance assessment}\label{sec:4.4.3}

Finally, the approximate average tracking and economic performance can be evaluated through integrating the joint normal probability density function (\ref{eq:4.4.2}) and the closed-loop performance functions (\ref{eq:4.4.3}), which are shown as Eq. (\ref{eq:4.4.4}).
\begin{equation}\label{eq:4.4.4}
  \begin{array}{l}
    {\bar J_{{\rm{ep}}}} \approx \int\limits_{{{\boldsymbol{x}}_{\rm{s}}} - (3\sim 5){\boldsymbol{\sigma }}}^{{{\boldsymbol{x}}_{\rm{s}}} + (3\sim 5){\boldsymbol{\sigma }}} {p({\boldsymbol{x}}_{\rm m}){J_{{\rm{ep}}}}({\boldsymbol{x}}_{\rm m})} {\rm{d}}{\boldsymbol{x}_{\rm m}} \\
    {\bar J_{{\rm{tp}}}} \approx \int\limits_{{{\boldsymbol{x}}_{\rm{s}}} - (3\sim 5){\boldsymbol{\sigma }}}^{{{\boldsymbol{x}}_{\rm{s}}} + (3\sim 5){\boldsymbol{\sigma }}} {p({\boldsymbol{x}}_{\rm m}){J_{{\rm{tp}}}}({\boldsymbol{x}}_{\rm m})} {\rm{d}}{\boldsymbol{x}_{\rm m}}
  \end{array}
\end{equation}
where ${\boldsymbol{\sigma}}$ denotes the standard deviation vector of ${\boldsymbol x}_{\rm m}$.

Theoretically, the performance assessment will be more accurate if the integral region is larger. However, the integral region cannot be too large because the calculated controller gain ${\boldsymbol K}^{\rm{MPC}}$ only works well around the given steady state. Thereby, the integral region can be set as $(3 \sim 5) {\boldsymbol\sigma}$, i.e. $\left[ {{{\boldsymbol{x}}_{\rm{s}}} - (3 \sim 5){\boldsymbol{\sigma }}, {{\boldsymbol{x}}_{\rm{s}}} + (3 \sim 5){\boldsymbol{\sigma }}} \right]$, which should be enough to cover more than 99.75\% states for a well-controlled steady-state operating system.

\section{Application to a chemical process example}\label{sec:5}

In this section, we apply the proposed procedure to the CSTR shown in Figure \ref{fig:3.0.1} to illustrate the applicability and effectiveness of the proposed approach. The model expression has been carefully explained in Section \ref{sec:3}. Four different cases are considered including Case 1 - $\boldsymbol x$ and $\boldsymbol u$ are both away from their boundaries, Case 2 - $\boldsymbol x$ locates on its boundaries and $\boldsymbol u$ is free (away from its boundaries), Case 3 - $\boldsymbol u$ locates on its boundaries and $\boldsymbol x$ is free, and Case 4 - other conditions.

\subsection{Case 1: $\boldsymbol x$ and $\boldsymbol u$ are both away from their boundaries}\label{sec:5.1}
In Case 1, we configure one scenario that the optimal $\boldsymbol x$ and $\boldsymbol u$ are both away from their boundaries, which is very common and desired in practical engineering. We are going to pre-evaluate the dynamic economic and tracking performance of EMPC and tracking MPC based on the proposed approach, which is helpful for us to choose an appropriate controller to achieve better performance. Simulation configuration has been described in detail by the first group experiment in Section \ref{sec:3}.

\subsubsection{Economic MPC}
Based on the above configuration, the optimal economic steady state is ${{\boldsymbol{x}}_{\rm{s}}} = [1.1601 ~615.7373]^{\rm T}$. The controller gain of EMPC obtained from sensitivity analysis is shown as below:
\begin{equation}\label{eq:5.1.5}
  {{\boldsymbol{K}}^{{\rm{EMPC}}}} = \left[ {\begin{array}{*{20}{c}}
  4.4428e\!+\!02 & 4.4232e\!+\!00\\
  1.9369e\!+\!07 & 1.8381e\!+\!05
  \end{array}} \right]
\end{equation}

\begin{figure}
    \centering
    \subfigure[State trajectories]{
      \label{subfig:5.1.1a} 
      \includegraphics[width=0.475\textwidth,angle=0]{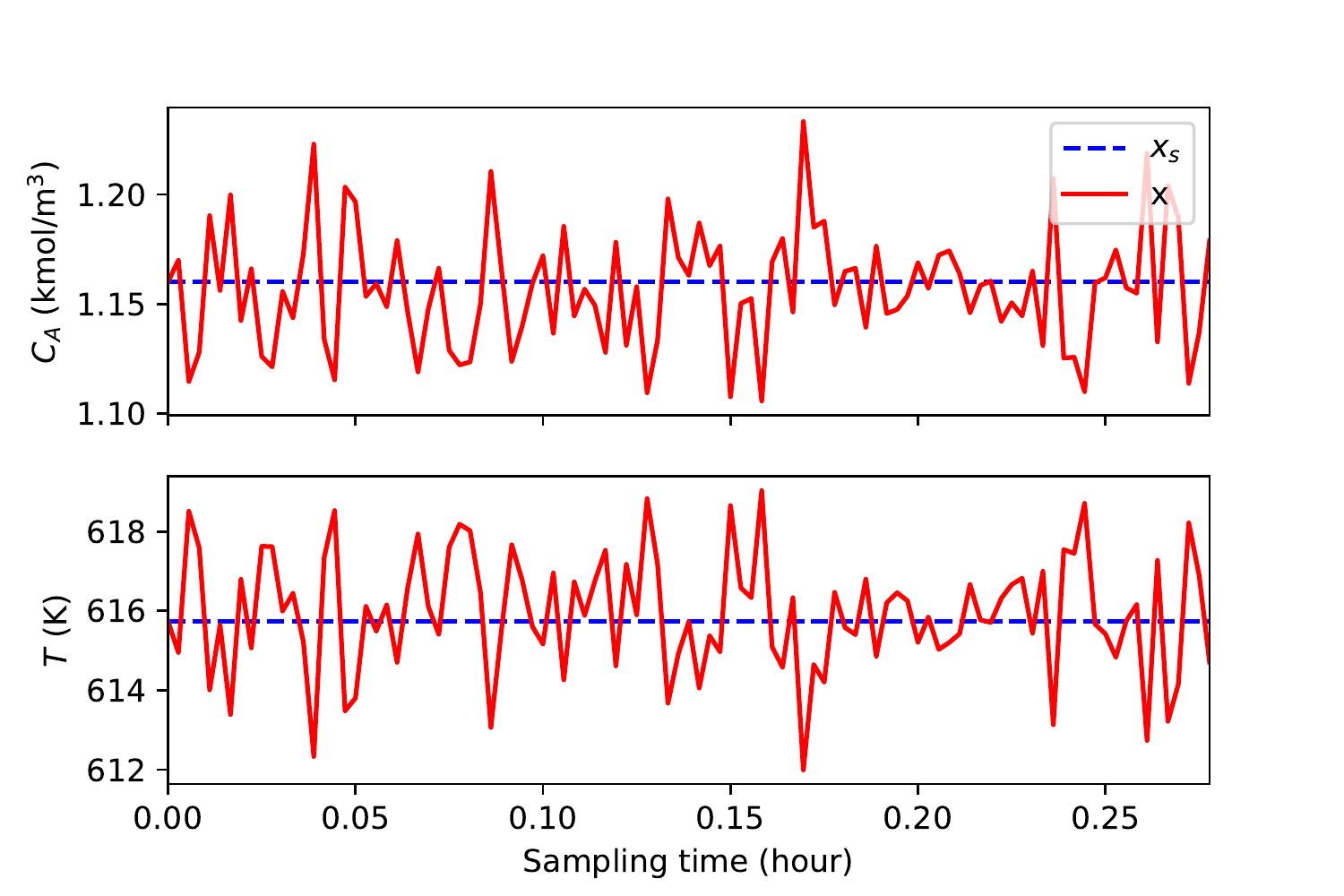}}
    \hspace{0in}
    \subfigure[Input trajectories]{
      \label{subfig:5.1.1b} 
      \includegraphics[width=0.475\textwidth,angle=0]{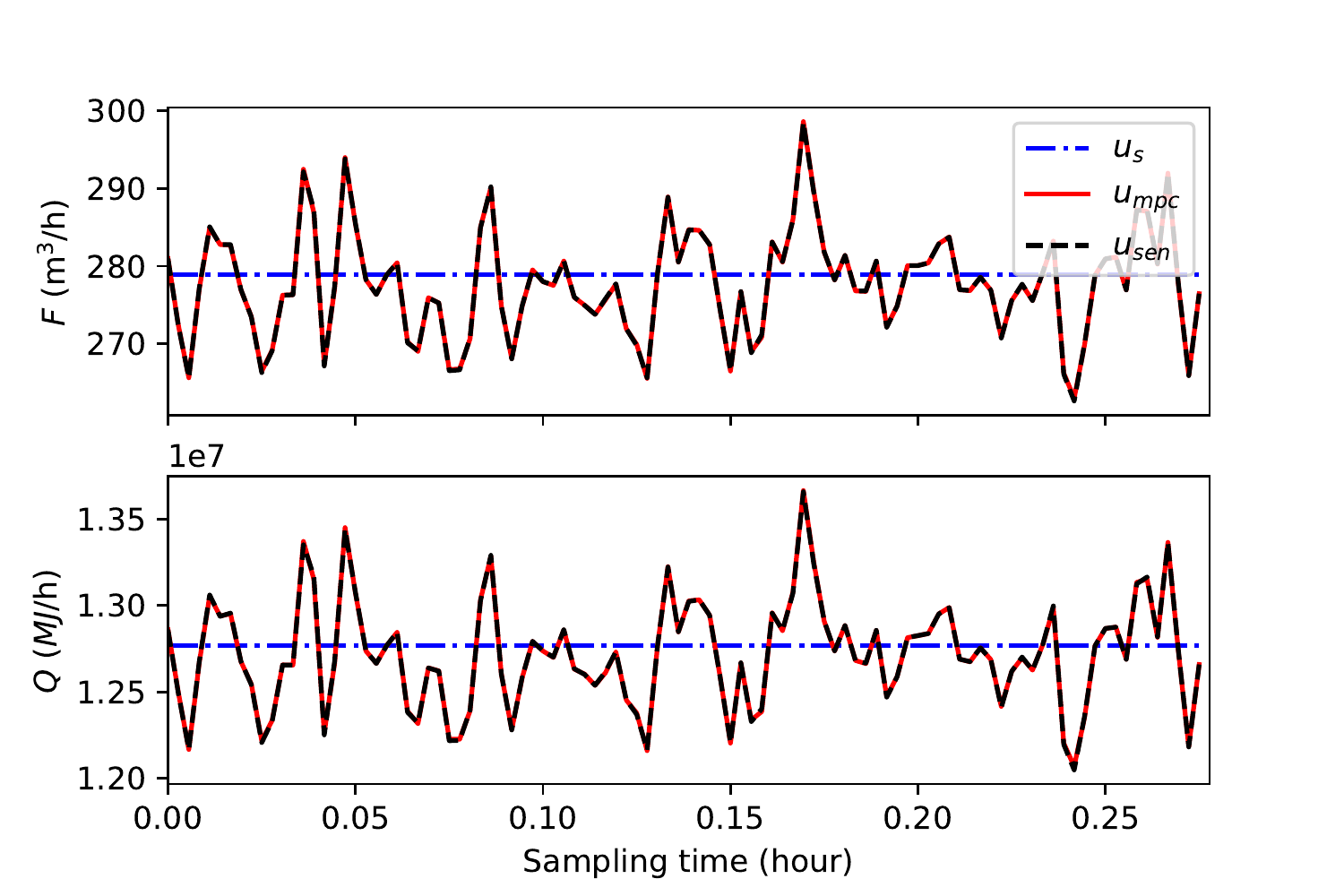}}
    \caption{State and input trajectories of EMPC in Case 1}\label{fig:5.1.1}
\end{figure}

Using EMPC and sensitivity analysis, we can get two groups input trajectories as Figure \ref{subfig:5.1.1b}based on the same state trajectory shown in Figure \ref{subfig:5.1.1a}, i.e. the optimal dynamic input trajectory generated by EMPC and the calculated input trajectory using sensitivity analysis. In Figure \ref{fig:5.1.1}, \textquotedblleft{$\rm x_s$}\textquotedblright\ and \textquotedblleft{$\rm u_s$}\textquotedblright\ denote the optimal state and input steady states, respectively; \textquotedblleft{$\rm x$}\textquotedblright\ is the actual state trajectory; \textquotedblleft{$\rm u_{mpc}$}\textquotedblright\ denotes the optimal input trajectory generated by EMPC; and \textquotedblleft{$\rm u_{sen}$}\textquotedblright\ denotes the calculated input trajectory using sensitivity analysis. It is shown that the \textquotedblleft{$\rm u_{mpc}$}\textquotedblright\ generated by EMPC and the \textquotedblleft{$\rm u_{sen}$}\textquotedblright\ from sensitivity analysis are basically the same. We can conclude that the controller gain obtained from sensitivity analysis is accurate enough.

Using the proposed approach, we can get the mean and variance of each variable. Table \ref{tab:5.1.1} shows the corresponding results, where $\mu$ denotes the mean, $\sigma^2$ is the variance, \textquotedblleft{Sim}\textquotedblright\ denotes the results from simulations, and \textquotedblleft{Cal}\textquotedblright\ denotes the results calculated by sensitivity evaluation. It is shown that the calculated results and the simulated results are almost exactly the same.

\begin{table}
    \caption{Mean and variance of each variable of EMPC in Case 1}\label{tab:5.1.1}
    \centering
    \begin{tabular}{llllllll}
      \hline
               &            &         $x_1$ &         $x_2$ &         $y_1$ &         $y_2$ &         $u_1$ &         $u_2$ \\
      \hline
           \multirow{2}{*}{$\mu$} &        Sim & 1.1603 & 615.7342 &  1.1603 & 615.7334 & 2.7894e+02 & 1.2774e+07 \\

               &        Cal & 1.1601 & 615.7373 & 1.1601 & 615.7373 & 2.7885e+02 & 1.2769e+07 \\
            \multirow{2}{*}{$\sigma^2$} &        Sim & 8.2736e-04 & 2.4467e+00 & 9.1822e-04 & 2.5323e+00 & 5.5399e+01 & 1.1233e+11 \\

               &        Cal & 8.3498e-04 & 2.4718e+00 & 9.2498e-04 & 2.5618e+00 & 5.5615e+01 & 1.1276e+11 \\
      \hline
    \end{tabular}
\end{table}

\begin{figure}
    \centering
    \subfigure[Economic performance surface]{
      \label{subfig:5.1.2a} 
      \includegraphics[width=0.475\textwidth,angle=0]{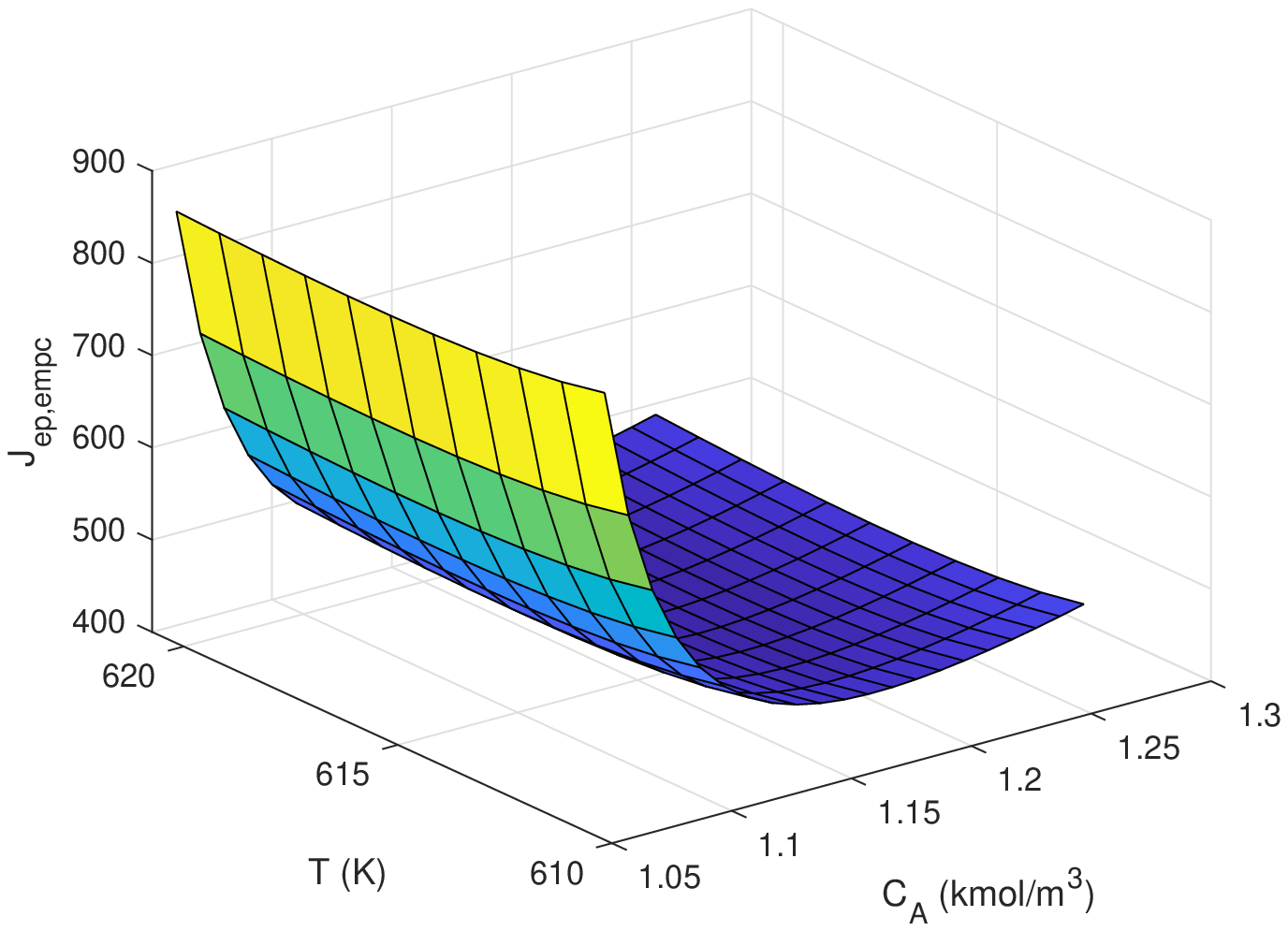}}
    \hspace{0in}
    \subfigure[Tracking performance surface]{
      \label{subfig:5.1.2b} 
      \includegraphics[width=0.475\textwidth,angle=0]{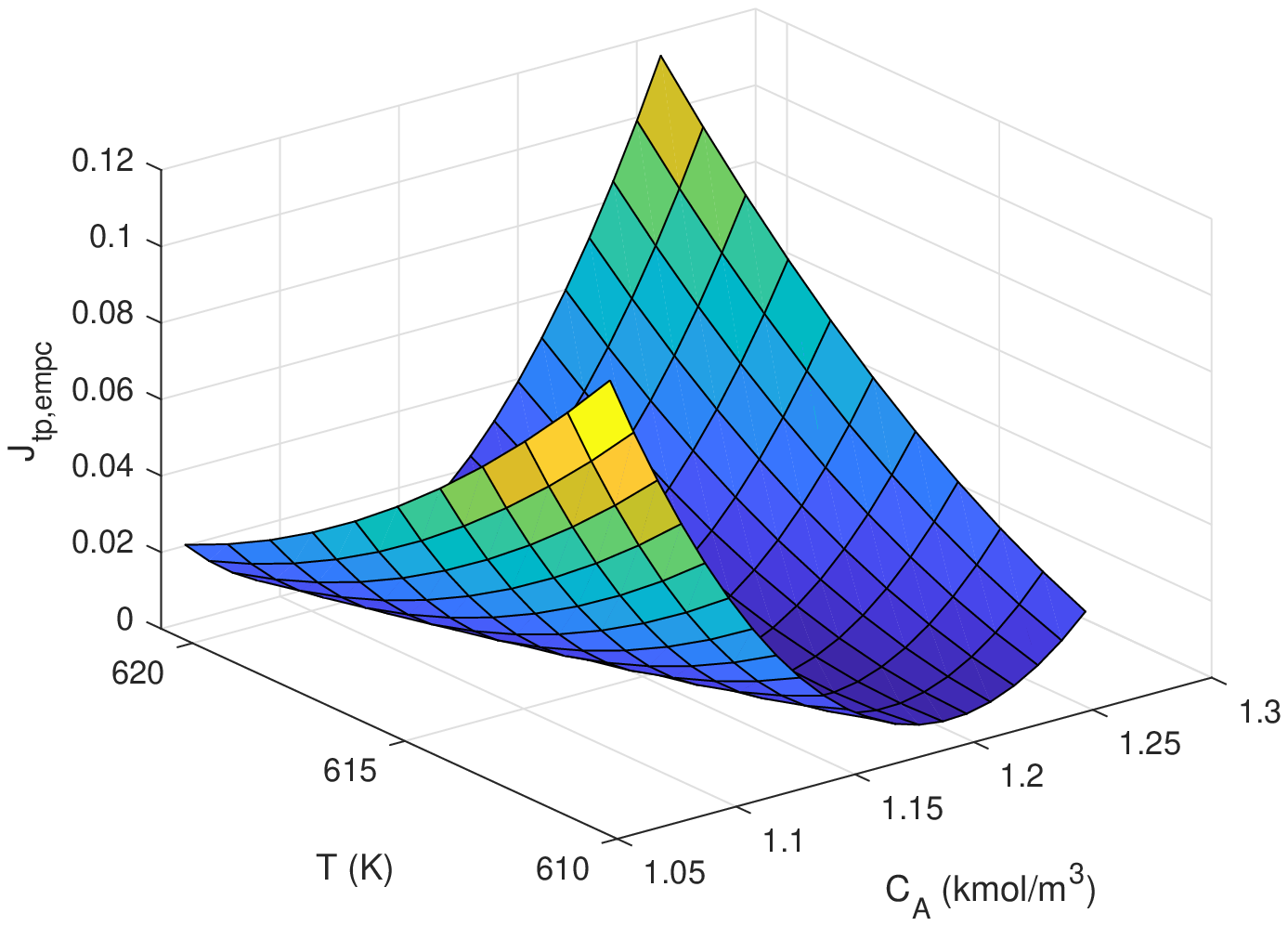}}
    \caption{Dynamic performance surfaces of EMPC in Case 1}\label{fig:5.1.2}
\end{figure}

Figure \ref{fig:5.1.2} shows the dynamic economic and tracking performance surfaces around the optimal steady state. It is shown that the optimal dynamic performance point locates around the optimal steady state. However, the dynamic performance surfaces may be not symmetric.

Suppose the two states satisfy joint normal distribution and their joint probability density function is as below:
\begin{equation}\label{eq:5.1.6}
  p({\boldsymbol x}_{\rm m}) = {\left( {2\pi {\sigma _{x_1}}{\sigma _{x_2}}\sqrt {1 - {\rho ^2}} } \right)^{ - 1}} e^{ { - \frac{1}{{2(1 - {\rho ^2})}}\left( {\frac{{{{({x_1} - {\mu _{x_1}})}^2}}}{{\sigma _{x_1}^2}} - \frac{{2\rho ({x_1} - {\mu _{x_1}})({x_2} - {\mu _{x_2}})}}{{{\sigma _{x_1}}{\sigma _{x_2}}}} + \frac{{{{({x_2} - {\mu _{x_2}})}^2}}}{{\sigma _{x_2}^2}}} \right)} }
\end{equation}
where $\mu_{x_i}$ denotes the mean of $x_i$, $\rho$ is the correlation coefficient.

Table \ref{tab:5.1.2} shows the average performance indices from simulations and calculations. To avoid the uncertainties of experiments, the simulation length is set very large to be $N_{\rm sim} = 100000$. The calculated average performance indices using the proposed method includes the results of integrating $3\boldsymbol\sigma$, $4\boldsymbol\sigma$, and $5\boldsymbol\sigma$ zones. It is shown that the calculated economic and tracking performance results are the same as the results of simulations, which reveals that the proposed approach works well in this situation.

\begin{table}
    \caption{Calculated and simulated performance indices of EMPC in Case 1}\label{tab:5.1.2}
    \centering
    \begin{tabular}{lllll}
      \hline
              &  & 3$\boldsymbol\sigma$ zone & 4$\boldsymbol\sigma$ zone & 5$\boldsymbol\sigma$ zone \\
      \hline
            \multirow{2}{*}{$\bar J_{\rm ep}$} &        Cal & 451.4201 & 453.9029 & 454.0342 \\

               &        Sim &  & 453.5738 &  \\
            \multirow{2}{*}{$\bar J_{\rm tp}$} &        Cal & 2.0345e-03 & 2.0831e-03 & 2.0475e-03 \\

               &        Sim &  & 2.0905e-03 &  \\
      \hline
    \end{tabular}
\end{table}

\subsubsection{Tracking MPC}

The controller gain of tracking MPC obtained from sensitivity analysis is shown as below:
\begin{equation}\label{eq:5.1.7}
  {{\boldsymbol{K}}^{{\rm{TMPC}}}} = \left[ {\begin{array}{*{20}{c}}
  2.8538e\!+\!02 & 2.8488e\!+\!00\\
  1.4720e\!+\!07 & 1.3795e\!+\!05
  \end{array}} \right]
\end{equation}

\begin{figure}
    \centering
    \subfigure[State trajectories]{
      \label{subfig:5.1.3a} 
      \includegraphics[width=0.475\textwidth,angle=0]{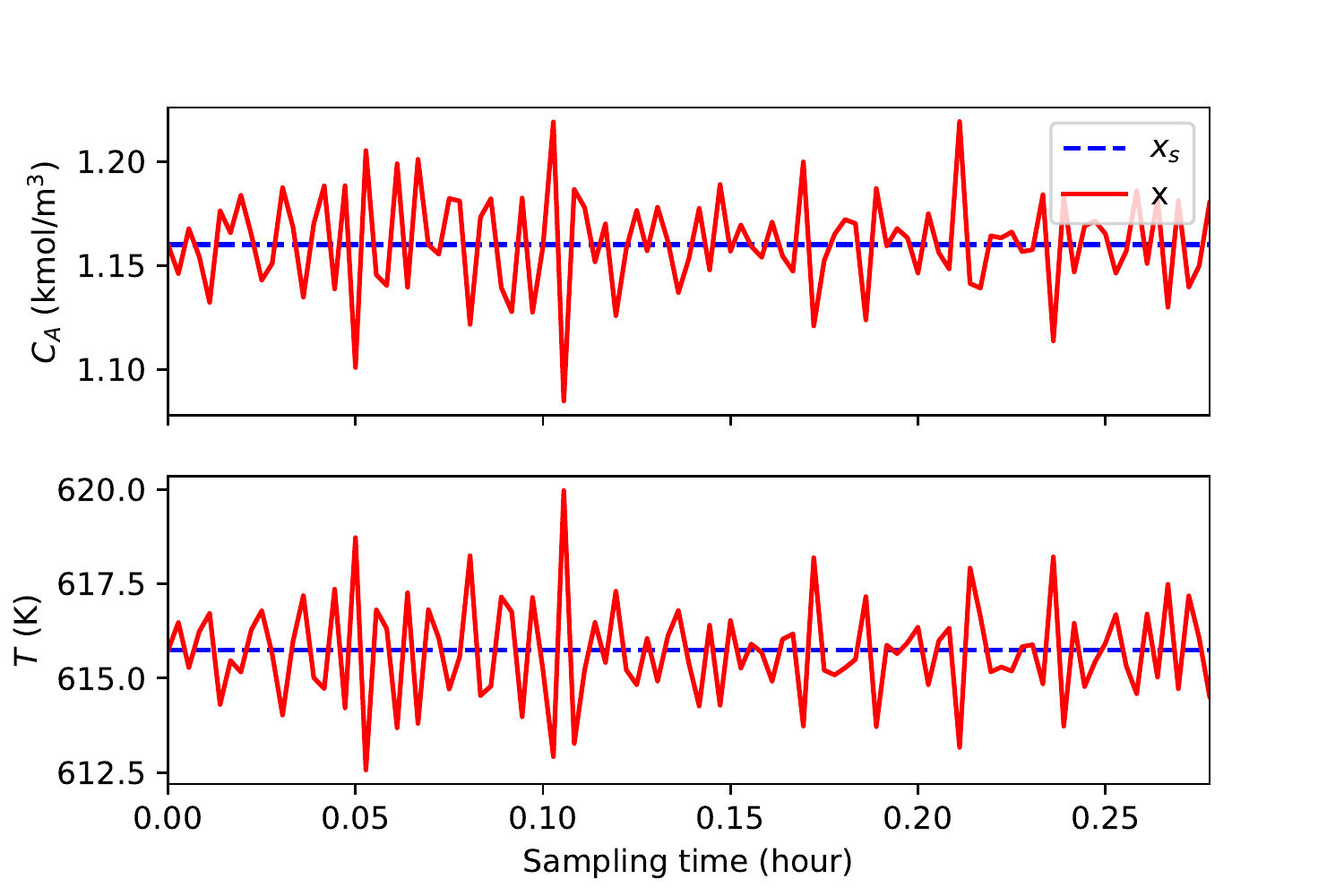}}
    \hspace{0in}
    \subfigure[Input trajectories]{
      \label{subfig:5.1.3b} 
      \includegraphics[width=0.475\textwidth,angle=0]{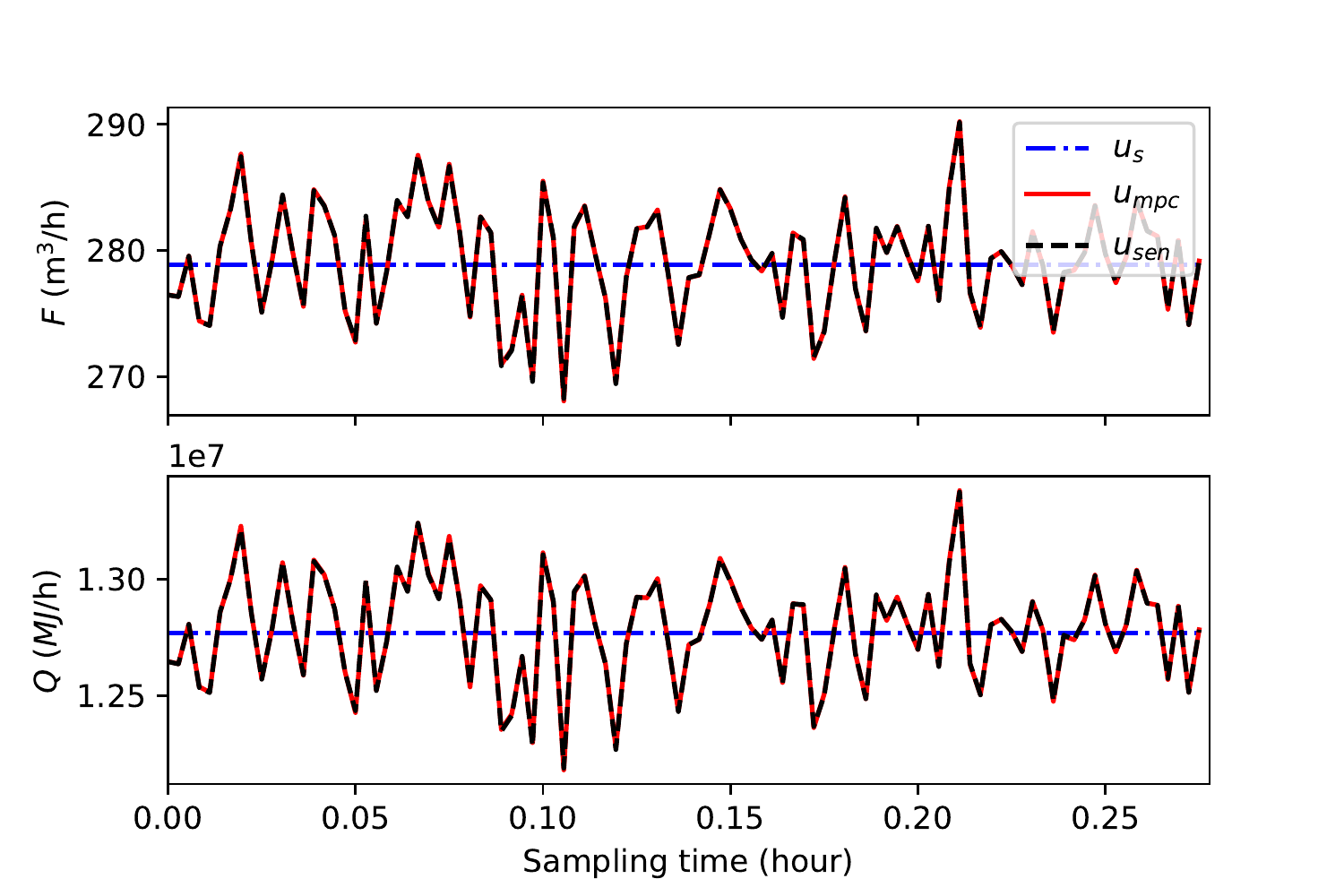}}
    \caption{State and input trajectories of tracking MPC in Case 1}\label{fig:5.1.3}
\end{figure}

\begin{figure}
    \centering
    \subfigure[Economic performance surface]{
      \label{subfig:5.1.4a} 
      \includegraphics[width=0.475\textwidth,angle=0]{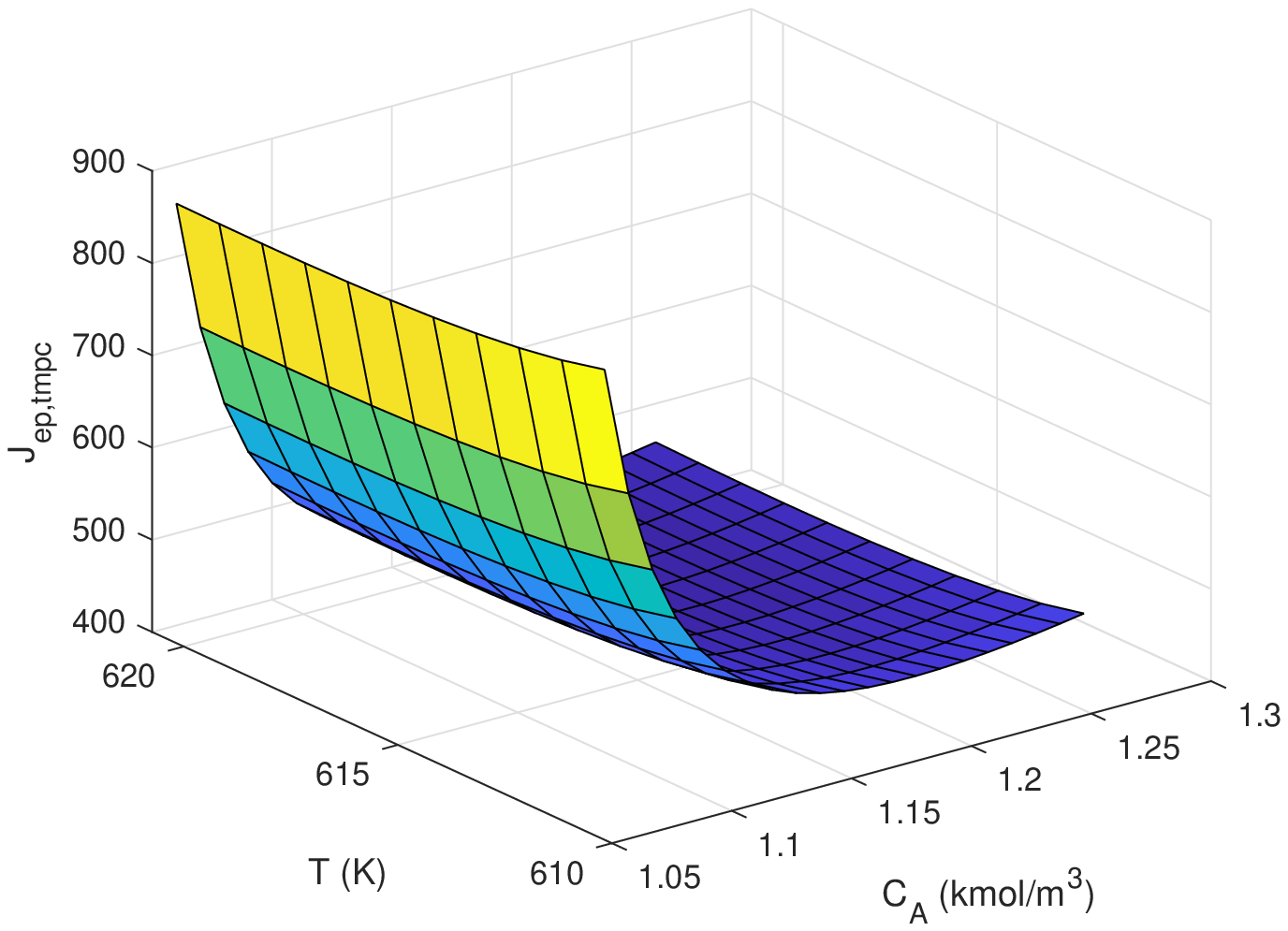}}
    \hspace{0in}
    \subfigure[Tracking performance surface]{
      \label{subfig:5.1.4b} 
      \includegraphics[width=0.475\textwidth,angle=0]{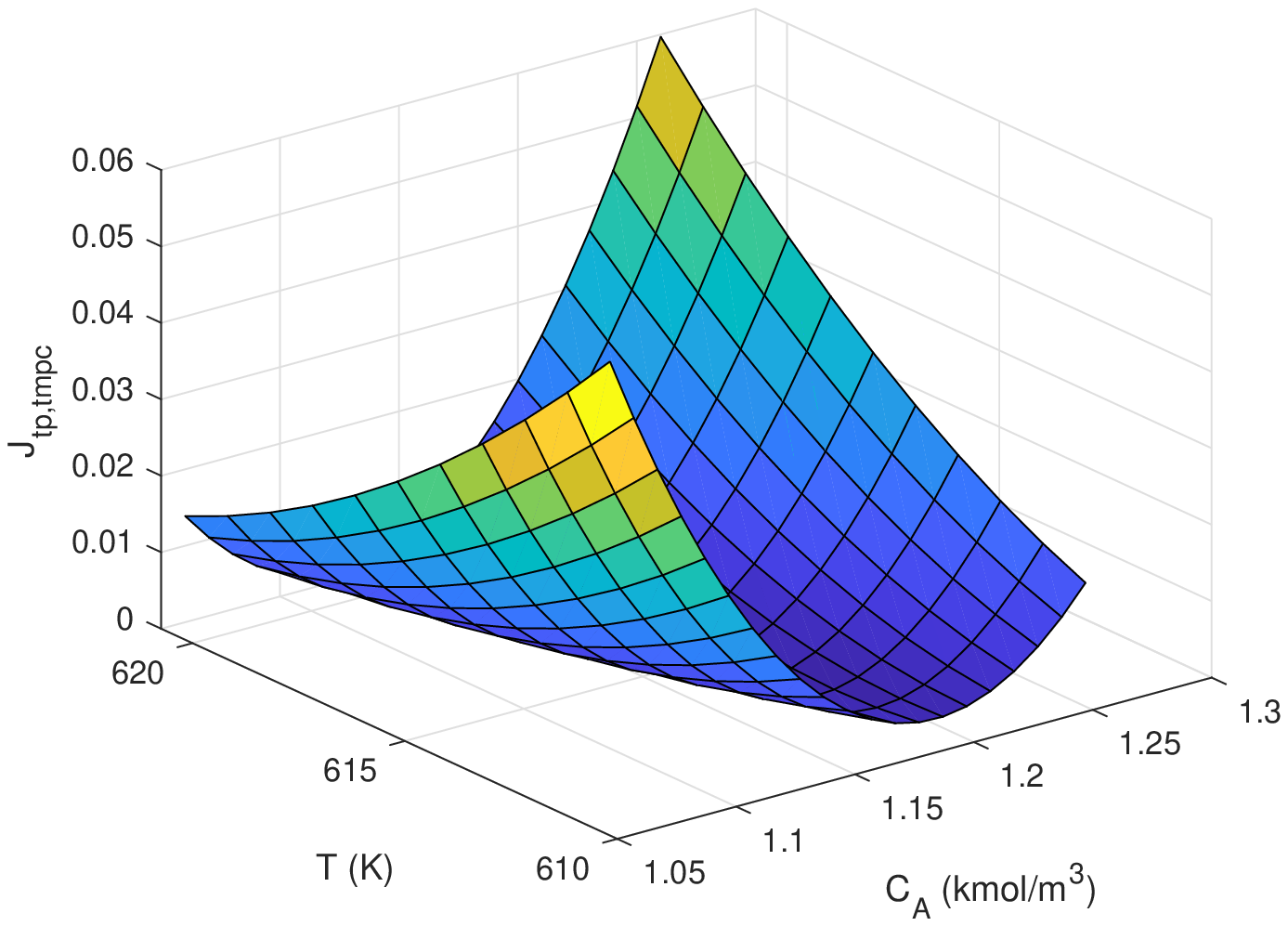}}
    \caption{Dynamic performance surfaces of tracking MPC in Case 1}\label{fig:5.1.4}
\end{figure}

The same results and conclusions as EMPC's can be attained from Figure \ref{fig:5.1.3}, Table \ref{tab:5.1.3}, Figure \ref{fig:5.1.4}, and Table \ref{tab:5.1.4}. They further illustrate the effectiveness of the proposed sensitivity-based dynamic performance assessment approach.

\begin{table}
    \caption{Mean and variance of each variable of tracking MPC in Case 1}\label{tab:5.1.3}
    \centering
    \begin{tabular}{llllllll}
      \hline
               &            &         $x_1$ &         $x_2$ &         $y_1$ &         $y_2$ &         $u_1$ &         $u_2$ \\
      \hline
           \multirow{2}{*}{$\mu$} &        Sim &  1.1601 & 615.7488 & 1.1601 & 615.7480 & 2.7890e+02 & 1.2773e+07 \\

               &        Cal & 1.1601 & 615.7373 & 1.1601 & 615.7373 & 2.7885e+02 & 1.2769e+07 \\
           \multirow{2}{*}{$\sigma^2$} &        Sim & 5.0130e-04 & 1.5292e+00 & 5.9141e-04 & 1.6164e+00 & 1.9444e+01 & 5.4403e+10 \\

               &        Cal & 5.0824e-04 & 1.5499e+00 & 5.9824e-04 & 1.6399e+00 & 1.9513e+01 & 5.4636e+10 \\
      \hline
    \end{tabular}
\end{table}

\begin{table}
    \caption{Calculated and simulated performance indices of tracking MPC in Case 1}\label{tab:5.1.4}
    \centering
    \begin{tabular}{lllll}
      \hline
                &  & 3$\boldsymbol\sigma$ zone & 4$\boldsymbol\sigma$ zone & 5$\boldsymbol\sigma$ zone \\
      \hline
            \multirow{2}{*}{$\bar J_{\rm ep}$} &        Cal & 446.6845 & 448.9983 & 449.0709 \\

               &        Sim &  & 448.9241 &  \\
            \multirow{2}{*}{$\bar J_{\rm tp}$} &        Cal & 1.0010e-03 & 1.0287e-03 & 1.0040e-03 \\

               &        Sim &  & 1.0275e-03 &  \\
      \hline
    \end{tabular}
\end{table}

\subsubsection{Performance comparison}
\noindent (1) Performance analysis of group 1

Compare Figure \ref{fig:5.1.2} with Figure \ref{fig:5.1.4}, the economic performance surfaces of EMPC and tracking MPC are very similar, while the tracking performance surfaces being different in consideration of the values of z-axis. The possible reason should be that EMPC adopts a nonsymmetric economic objective while tracking MPC using a symmetric quadratic tracking objective  so that tracking MPC will be more sensitive to the variations of states. The differences can also be verified by Table \ref{tab:5.1.1} and Table \ref{tab:5.1.3}. Meanwhile, Table \ref{tab:5.1.1} and Table \ref{tab:5.1.3} show that the variances of variables in EMPC and tracking MPC are different, which makes their economic performance be different. Table \ref{tab:5.1.2} and Table \ref{tab:5.1.4} reveal that tracking MPC has 1\% better economic performance than EMPC in this group experiments.

\noindent (2) Performance comparison of group 1 and group 2

In group 2 of Section \ref{sec:3}, the optimal economic steady state is ${\boldsymbol x}_{\rm s} = [2.7284 ~631.7380]^{\rm T}$ and the average performance indices are shown in Table \ref{tab:5.1.5} and Table \ref{tab:5.1.6}. In this group experiments, the economic performance of EMPC is better than tracking MPC (about 1\%). Compare them with Table \ref{tab:5.1.2} and Table \ref{tab:5.1.4}, it is shown that which controller can achieve better performance is uncertain in practical processes. However, the proposed dynamic performance assessment approach can provide accurate pre-evaluation on the average performance of EMPC and tracking MPC, which can be used to choose the most appropriate controller when engineers design controllers for given systems.

\begin{table}
    \caption{Calculated and simulated performance indices of EMPC in group 2 of Case 1}\label{tab:5.1.5}
    \centering
    \begin{tabular}{lllll}
      \hline
                &  & 3$\boldsymbol\sigma$ zone & 4$\boldsymbol\sigma$ zone & 5$\boldsymbol\sigma$ zone \\
      \hline
            \multirow{2}{*}{$\bar J_{\rm ep}$} &        Cal & 197.9753 & 198.9069 & 198.9312 \\

               &        Sim &  & 199.8444 &  \\
            \multirow{2}{*}{$\bar J_{\rm tp}$} &        Cal & 6.7851e-03 & 7.0437e-03 & 7.0418e-03 \\

               &        Sim &  & 7.1578e-03 &  \\
      \hline
    \end{tabular}
\end{table}

\begin{table}
    \caption{Calculated and simulated performance indices of tracking MPC in group 2 of Case 1}\label{tab:5.1.6}
    \centering
    \begin{tabular}{lllll}
      \hline
                &  & 3$\boldsymbol\sigma$ zone & 4$\boldsymbol\sigma$ zone & 5$\boldsymbol\sigma$ zone \\
      \hline
            \multirow{2}{*}{$\bar J_{\rm ep}$} &        Cal & 200.6039 & 201.6109 & 201.6389 \\

               &        Sim &  & 202.5862 &  \\
            \multirow{2}{*}{$\bar J_{\rm tp}$} &        Cal & 6.5879e-03 & 6.7543e-03 & 6.7537e-03 \\

               &        Sim &  & 6.7702e-03 &  \\
      \hline
    \end{tabular}
\end{table}

\subsection{Case 2: $\boldsymbol x$ locates on its boundaries and $\boldsymbol u$ is free}\label{sec:5.2}
In Case 2, we configure one scenario that $\boldsymbol x$ locates on its boundaries and $\boldsymbol u$ is free (away from its boundaries). In this scenario, violation of boundaries and offset of steady states are very common in tracking MPC and EMPC, which will be presented in the follows. Based on the proposed performance assessment approach, we can pre-evaluate the violation degree and pre-set boundary moving or target moving to guarantee the stability of controlled processes.

The economic objective is set as below:
\begin{equation}\label{eq:5.2.1}
  {J_{\rm{ep}}} = {{x_1} + {x_2}}
\end{equation}

The constraints considered in the experiments are shown as follows:
\begin{equation}\label{eq:5.2.2}
      1 \le {x_1} \le 3, ~500 \le {x_2} \le 1000
\end{equation}

The noise added on states and measurements is shown as follows:
\begin{equation}\label{eq:5.2.3}
  \begin{array}{l}
    {w_1}\sim{\rm N}(0,0.0001),{v_1}\sim{\rm N}(0,0.1)\\
    {w_2}\sim{\rm N}(0,0.0004),{v_2}\sim{\rm N}(0,0.4)
  \end{array}
\end{equation}

\subsubsection{Violation and offset}
\noindent (1) Problem description

\begin{figure}
    \centering
    \subfigure[Violation phenomenon]{
      \label{subfig:5.2.2a} 
      \includegraphics[width=0.475\textwidth,angle=0]{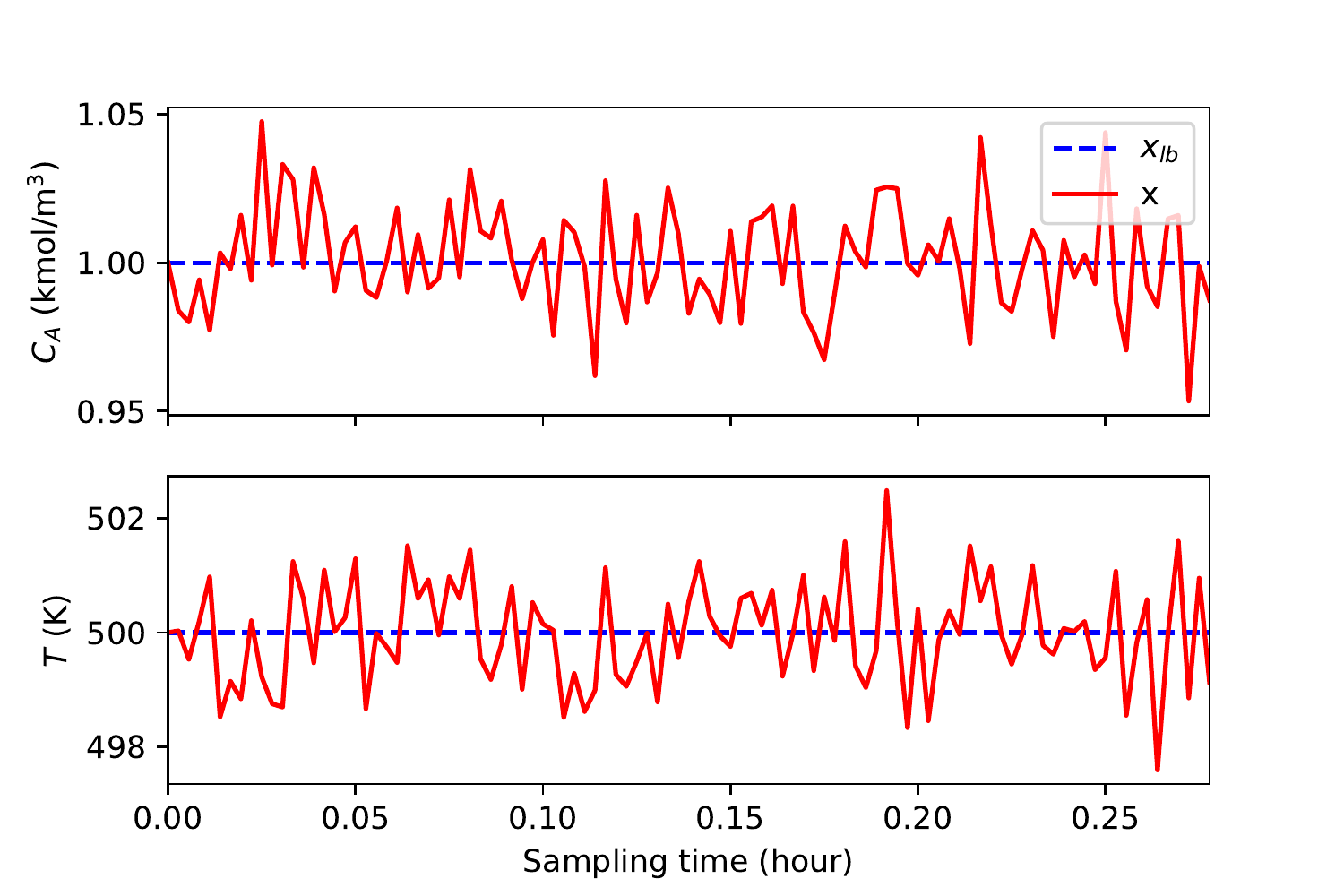}}
    \hspace{0in}
    \subfigure[Offset phenomenon]{
      \label{subfig:5.2.2b} 
      \includegraphics[width=0.475\textwidth,angle=0]{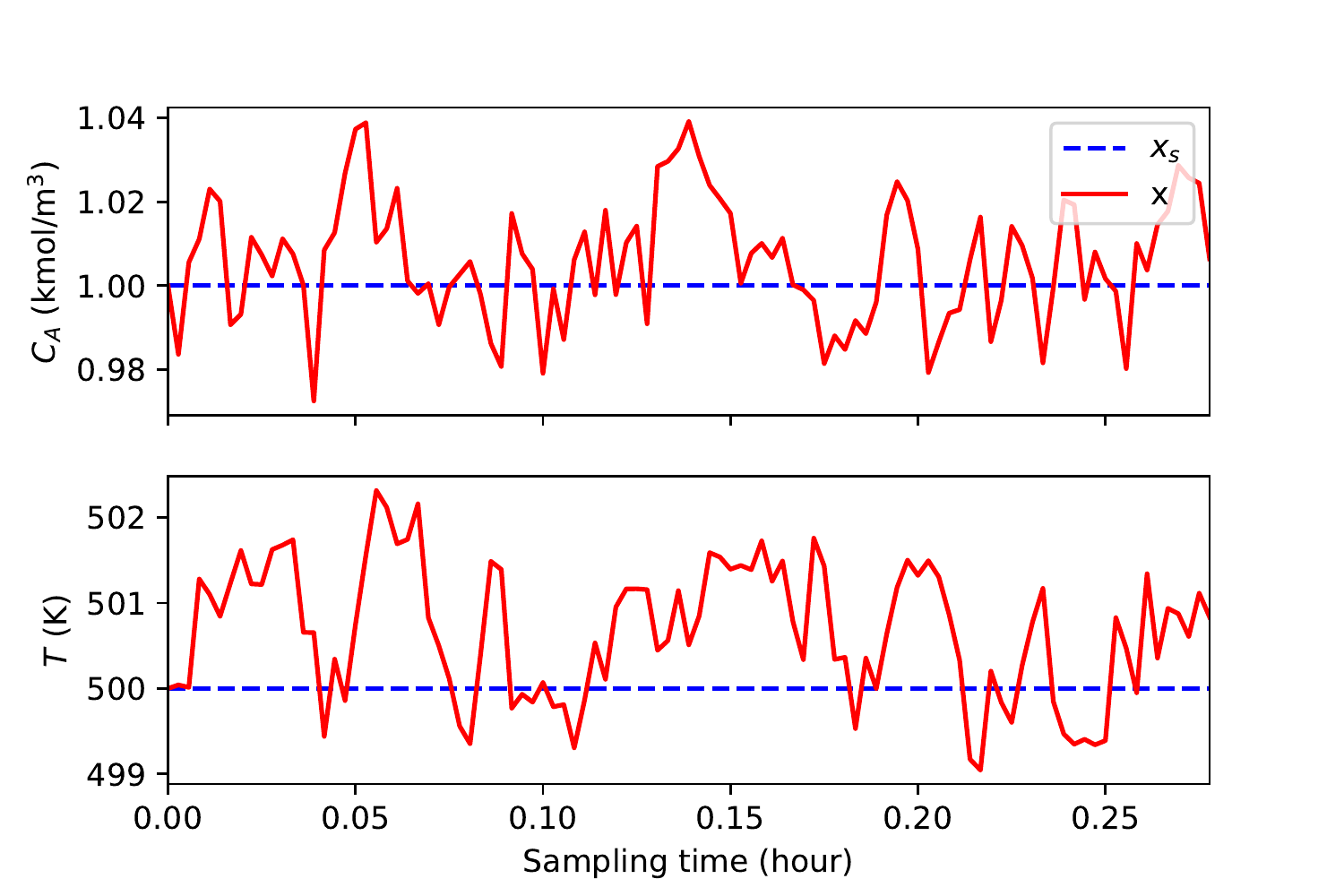}}
    \caption{Violation and offset phenomenons of state variables}\label{fig:5.2.2}
\end{figure}

In the EMPC and tracking MPC controlled systems, when $\boldsymbol x$ is close to its boundaries, the actual state values will violate their boundaries in the presence of noise, which is harmful to the safety and stability of controlled systems. Figure \ref{subfig:5.2.2a} gives a simple example to show this problem, in which \textquotedblleft{$\rm x_{lb}$}\textquotedblright\ denotes the lower boundary of $\boldsymbol x$. Except this, there is another problem called offset shown in Figure \ref{subfig:5.2.2b}, which means that the actual states will diverge from the given steady state ${{\boldsymbol{x}}_{\rm{s}}}$. It may be caused by the punishment of both ${{\boldsymbol{x}} - {{\boldsymbol{x}}_{\rm{s}}}}$ and ${{\boldsymbol{u}} - {{\boldsymbol{u}}_{\rm{s}}}}$ in tracking MPC. The violation problem exists in both EMPC and tracking MPC, while the offset problem existing in only tracking MPC for the reason that EMPC is one kind of optimal dynamic performance searching not a given target tracking.

\noindent (2) Reasonable solutions

\textcircled{1} Move the boundaries (Boundary moving)

If there is not offset, moving the boundaries of $\boldsymbol x$ (i.e. backoff) will be a good method to avoid the violation while ensuring minimum economy loss. Therefore, it is a suitable way to deal with the violation in EMPC. However, this method works not so well with tracking MPC because the existence of offset in tracking MPC will make the states diverge from the adjusted boundaries again and lead to more economy loss.

\textcircled{2} Move the optimal steady state (Target moving)

When the optimal steady state locates on the boundaries, there may co-exist violation and offset problems in tracking MPC. In this situation, moving the optimal steady state away from its boundaries will be a good method to solve these problems. After target moving, the reformulated optimization problem will be the same as Case 1 in Section \ref{sec:5.1}. Target moving is a suitable way to deal with the violation and offset in tracking MPC but not suitable for EMPC due to the lack of given steady states.

\subsubsection{Economic MPC}\label{sec:5.2.2}
The optimal steady state locates on the original lower boundary shown as below:
\begin{equation}\label{eq:5.2.4}
  {{\boldsymbol{x}}_{\rm{s}}} = {\boldsymbol{x}}_{{\rm{ori}}}^{{\rm{lb}}} = [1 ~500]^{\rm T}
\end{equation}

The controller gain of EMPC obtained from sensitivity analysis is as below:
\begin{equation}\label{eq:5.2.5}
  {{\boldsymbol{K}}^{{\rm{EMPC}}}} = \left[ {\begin{array}{*{20}{c}}
  -9.5472e\!+\!01 & 4.8640e\!-\!01\\
  -5.5836e\!+\!06 & -7.0123e\!+\!04
  \end{array}} \right]
\end{equation}

The variances of states calculated by the proposed method are shown as below:
\begin{equation}\label{eq:5.2.6}
  {\boldsymbol{\sigma}}^2 = [2.8039e\!-\!04 ~~~6.7546e\!-\!01]^{\rm T}
\end{equation}

The new lower boundary obtained through moving $3\boldsymbol{\sigma}$ distance is as below:
\begin{equation}\label{eq:5.2.7}
  \begin{array}{l}
  {\boldsymbol{x}}_{{\rm{new}}}^{{\rm{lb}}} = {\boldsymbol{x}}_{{\rm{ori}}}^{{\rm{lb}}} + 3{\boldsymbol{\sigma }} = [1.0502 ~502.4656]^{\rm T}
  \end{array}
\end{equation}

\begin{figure}
    \centering
    \subfigure[State trajectories]{
      \label{subfig:5.2.3a} 
      \includegraphics[width=0.475\textwidth,angle=0]{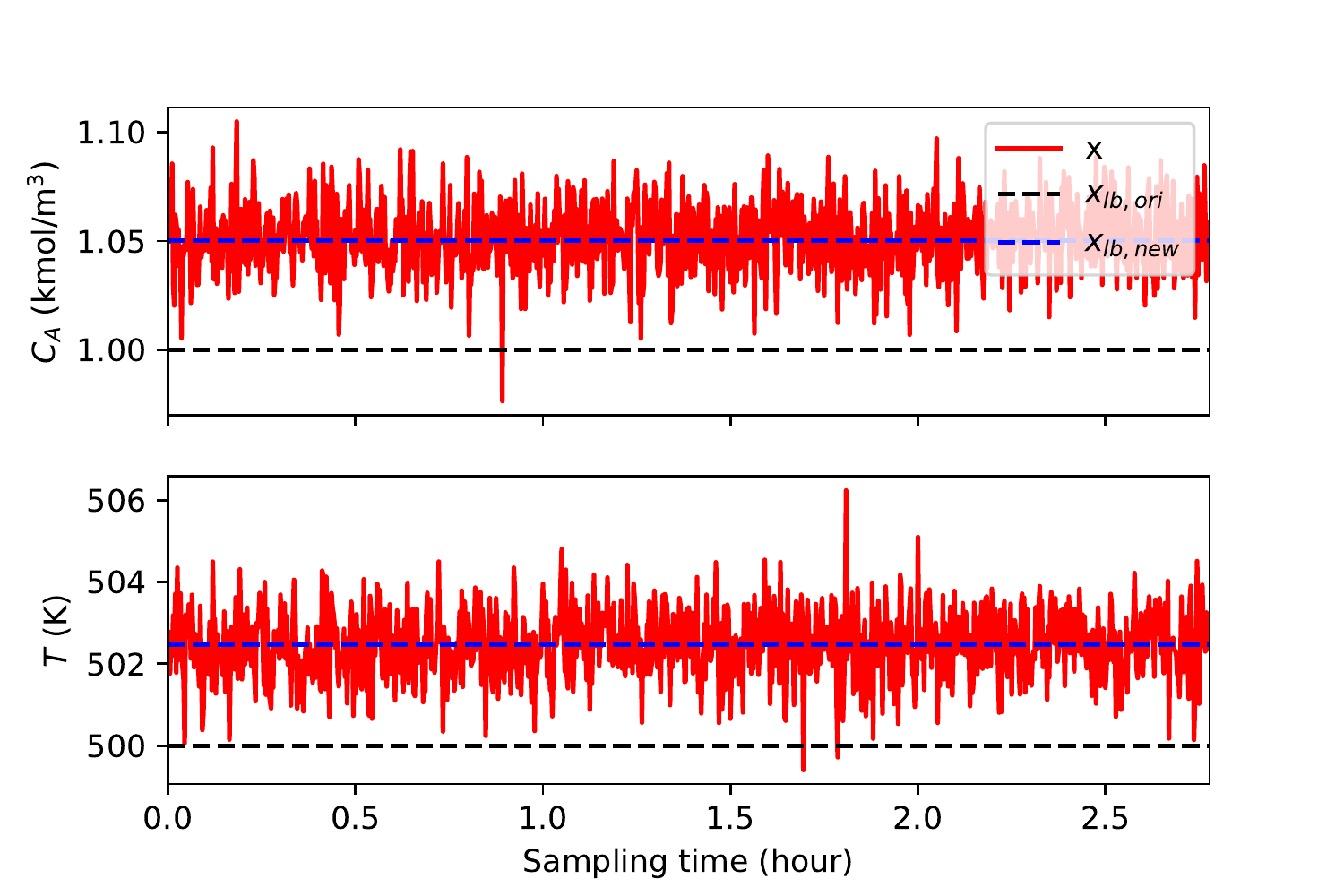}}
    \hspace{0in}
    \subfigure[Input trajectories]{
      \label{subfig:5.2.3b} 
      \includegraphics[width=0.475\textwidth,angle=0]{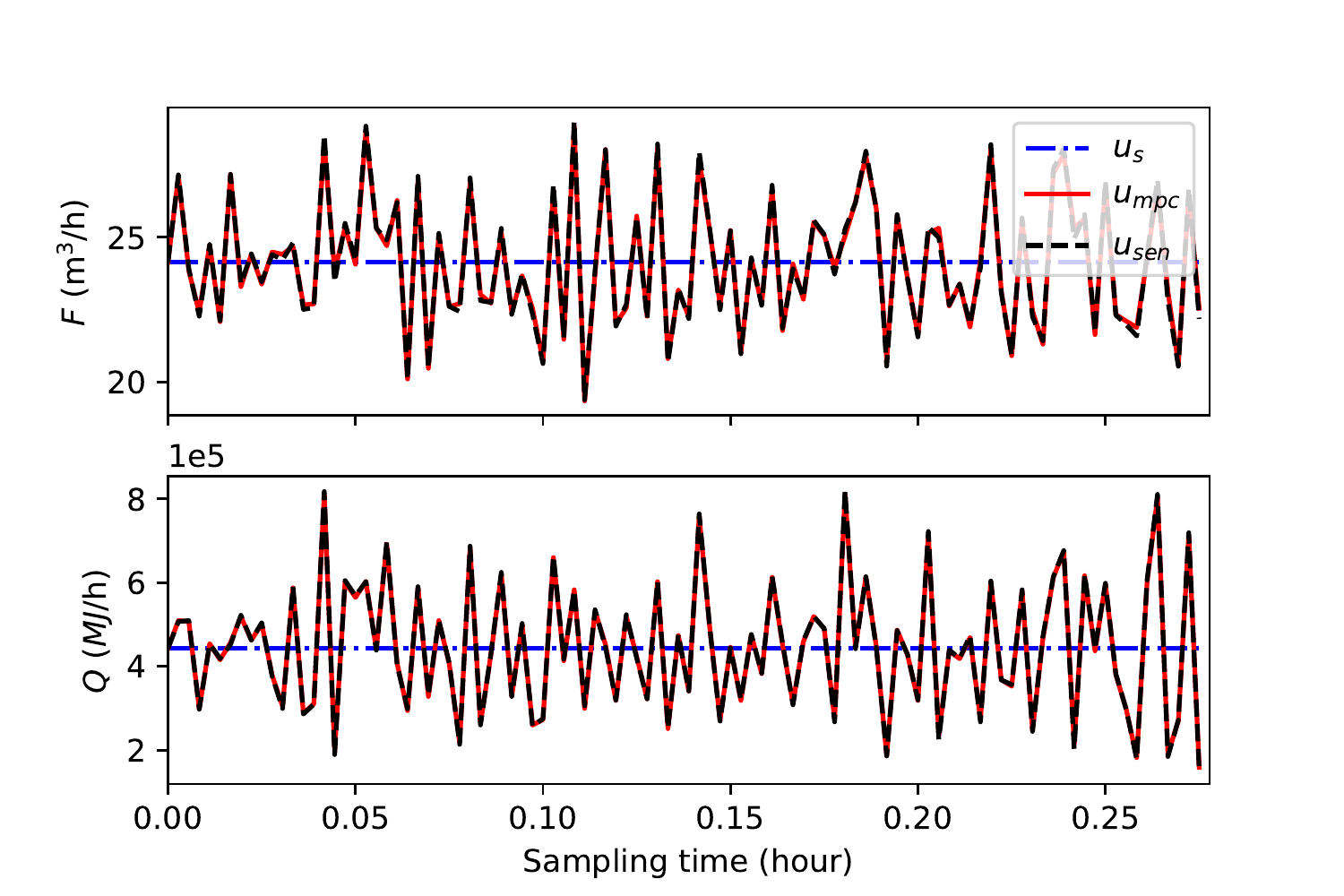}}
    \caption{Simulation results after boundary moving of EMPC in Case 2}\label{fig:5.2.3}
\end{figure}

The simulation results after boundary moving are shown in Figure \ref{fig:5.2.3}, in which \textquotedblleft{$\rm x_{lb,ori}$}\textquotedblright\ and \textquotedblleft{$\rm x_{lb,new}$}\textquotedblright\ denote the original and new lower boundaries of $\boldsymbol x$ respectively. To show the violation of $\boldsymbol x$, we plot more than two hours results in Figure \ref{subfig:5.2.3a}. It is shown the percentages of the data that violates the original lower boundary are $[0.1036\% ~0.1978\%]^{\rm T}$, which are close to the theoretical values $[0.125\% ~0.125\%]^{\rm T}$. Therefore, the sensitivity-based boundary moving works well for reducing the risk of violation in EMPC.

\subsubsection{Tracking MPC}\label{sec:5.2.3}
The original given steady state and lower boundary are shown as below:
\begin{equation}\label{eq:5.2.8}
  {{\boldsymbol{x}}_{{\rm{s,ori}}}} = {{\boldsymbol{x}}^{{\rm{lb}}}} = [1 ~500]^{\rm T}
\end{equation}

The reformulated problem after target moving will be a ``Case 1'' problem shown in Section \ref{sec:5.1}. The actual variances of states at the new adjusted steady state is unknown because the adjusted steady state cannot be determined before knowing its variances. However, the variances at the original steady state can be used to replace the variances at the new adjusted steady state because the adjusted steady state will not be far away from the original steady state. It is worthy noting that the active constraint flags should be set as ${{\boldsymbol{I}}_{\boldsymbol{x}}} = {\boldsymbol{0}},{{\boldsymbol{I}}_{\boldsymbol{u}}} = {\boldsymbol{0}}$ when we calculate the controller gain of tracking MPC.

The controller gain of tracking MPC obtained from sensitivity analysis is as below:
\begin{equation}\label{eq:5.2.9}
  {{\boldsymbol{K}}^{{\rm{TMPC}}}} = \left[ {\begin{array}{*{20}{c}}
  1.3458e\!-\!01 & 1.5024e\!-\!01\\
  -4.1946e\!+\!04 & -1.2903e\!+\!03
  \end{array}} \right]
\end{equation}

The variances of states calculated by sensitivity analysis are as follows:
\begin{equation}\label{eq:5.2.10}
  {\boldsymbol{\sigma}}^2 = [2.4150e\!-\!04 ~~~2.0429e\!+\!00]^{\rm T}
\end{equation}

The new steady state obtained through moving $3\boldsymbol{\sigma}$ distance is as below:
\begin{equation}\label{eq:5.2.11}
  {{\boldsymbol{x}}_{{\rm{s,new}}}} = {{\boldsymbol{x}}_{{\rm{s,ori}}}} + 3{\boldsymbol{\sigma }} = [1.0466 ~504.2879]^{\rm T}
\end{equation}

\begin{figure}
    \centering
    \subfigure[State trajectories]{
      \label{subfig:5.2.4a} 
      \includegraphics[width=0.475\textwidth,angle=0]{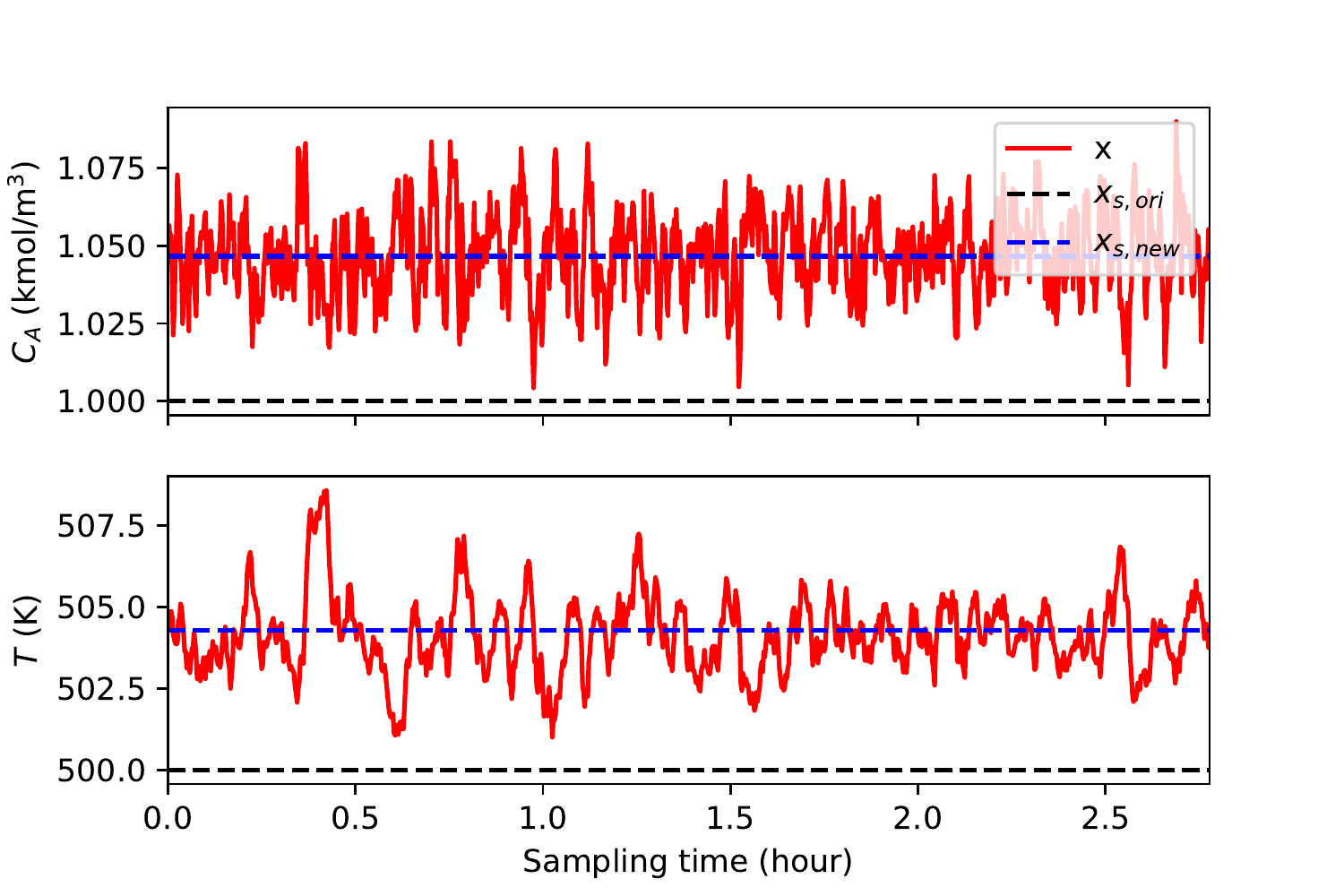}}
    \hspace{0in}
    \subfigure[Input trajectories]{
      \label{subfig:5.2.4b} 
      \includegraphics[width=0.475\textwidth,angle=0]{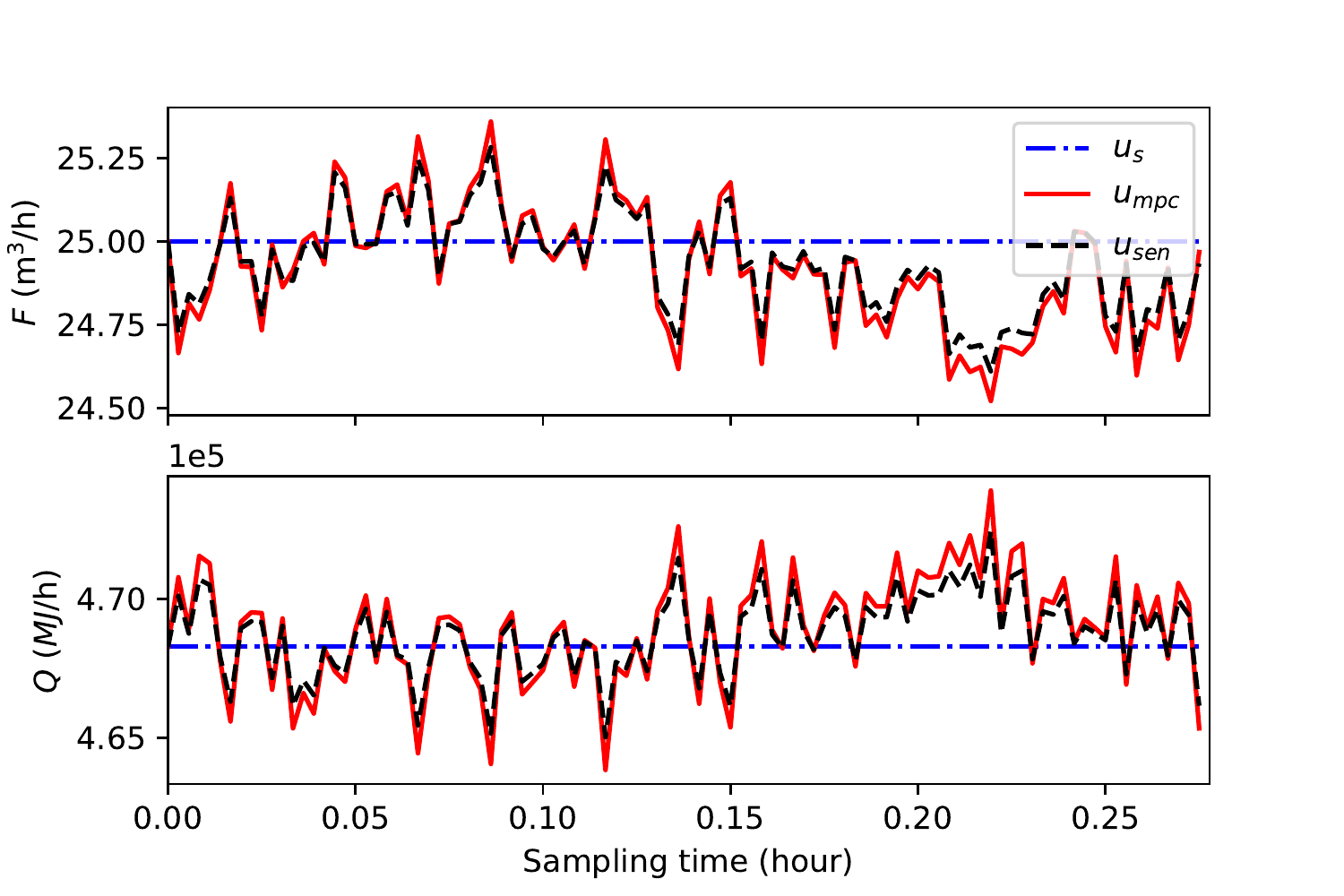}}
    \caption{Simulation results after target moving of TMPC in Case 2}\label{fig:5.2.4}
\end{figure}

The simulation results after target moving are shown in Figure \ref{fig:5.2.4}, in which \textquotedblleft{$\rm x_{s,ori}$}\textquotedblright\ and \textquotedblleft{$\rm x_{s,new}$}\textquotedblright\ denote the original and new steady states of $\boldsymbol x$ respectively. It is shown the percentages of the data that violates the original lower boundary (${\boldsymbol x}^{\rm lb} = {\boldsymbol x}_{\rm s,ori}$) are $[0.0552\% ~0.0126\%]^{\rm T}$, which shows the proposed sensitivity-based target moving guarantees as little data as possible crosses the original boundaries. The differences between ${\boldsymbol u}_{\rm mpc}$ and ${\boldsymbol u}_{\rm sen}$ may result from the controller gain approximation for using ${\boldsymbol x}_{\rm s,ori}$ instead of ${\boldsymbol x}_{\rm s,new}$. In conclusion, the sensitivity-based target moving works well for reducing the risk of violation and offset in tracking MPC.

\subsubsection{Performance assessment}

Table \ref{tab:5.2.1} shows the economic performance loss resulted from boundary moving and target moving. The economy loss of EMPC is about 0.50\%. Both the simulation results and the proposed assessment method give the same conclusion. The economy loss of tracking MPC is about 0.87\%. The results illustrate that the proposed method could provide accurate pre-evaluation about the possible performance loss when boundary moving and target moving are applied in EMPC and tracking MPC.

\begin{table}
    \caption{Economic performance loss resulted from boundary moving and target moving}\label{tab:5.2.1}
    \centering
    \begin{tabular}{lllll}
      \hline
               &            &    Ideal $\bar J_{\rm{ep}}$ & Actual $\bar J_{\rm{ep}}$ & Loss of $\bar J_{\rm{ep}}$ \\
      \hline
          \multirow{2}{*}{Economic MPC} &        Cal &        501 & 503.5158 & 2.5158 (0.50\%) \\

               &        Sim &        501 & 503.5214 & 2.5214 (0.50\%) \\
          \multirow{2}{*}{Tracking MPC} &        Cal &        501 & 505.3345 & 4.3345 (0.87\%) \\

               &        Sim &        501 & 505.2858 & 4.2858 (0.86\%) \\
      \hline
    \end{tabular}
\end{table}

Another group of experiments are conducted as a comparison, in which the original lower boundary of $\boldsymbol x$ is changed to ${{\boldsymbol{x}}^{{\rm{lb}}}} = [1 ~600]^{\rm T}$. Table \ref{tab:5.2.2} shows the economic performance loss in the new experiments. Compare Table \ref{tab:5.2.2} with Table \ref{tab:5.2.1}, it is shown that which controller can achieve better performance is uncertain after boundary moving and target moving. But the proposed approach can tell us how much we need moving the boundaries or targets for stable operation and which controller can achieve less performance loss.

\begin{table}
    \caption{Economic performance loss resulted from boundary moving and target moving in group 2 of Case 2}\label{tab:5.2.2}
    \centering
    \begin{tabular}{lllll}
      \hline
               &            &    Ideal $\bar J_{\rm{ep}}$ & Actual $\bar J_{\rm{ep}}$ & Loss of $\bar J_{\rm{ep}}$ \\
      \hline
          \multirow{2}{*}{Economic MPC} &        Cal &        601 & 608.0289 & 7.0289 (1.17\%) \\

               &        Sim &        601 & 608.0316 & 7.0316 (1.17\%) \\
          \multirow{2}{*}{Tracking MPC} &        Cal &        601 & 605.5987 & 4.5987 (0.77\%) \\

               &        Sim &        601 & 605.5677 & 4.5677 (0.76\%) \\
      \hline
    \end{tabular}
\end{table}

\subsection{Case 3: $\boldsymbol x$ is free and $\boldsymbol u$ locates on its boundaries}\label{sec:5.3}
In Case 3, we configure one scenario that $\boldsymbol x$ is free (away from its boundaries) and $\boldsymbol u$ locates on its boundaries. In this scenario, Gaussian noise will influence the controller gain of tracking MPC due to the deficiency of control freedom caused by the active constraints of $\boldsymbol u$. This is not desired and expected to be avoided in process control.

The economic objective is shown as below \cite{li2016application}:
\begin{equation}\label{eq:5.3.1}
  J_{\rm ep} = - 1700000{u_1}({C_{A0}} - {x_1}) + {u_2}
\end{equation}

The constraints considered in the experiments are shown as follows:
\begin{equation}\label{eq:5.3.2}
    0 \le {u_1} \le 10,~ - 200000 \le {u_2} \le 200000
\end{equation}

The noise added on states and measurements is shown as follows:
\begin{equation}\label{eq:5.3.3}
  \begin{array}{l}
    {w_1}\sim{\rm N}(0,0.00001),{v_1}\sim{\rm N}(0,0.01)\\
    {w_2}\sim{\rm N}(0,0.00004),{v_2}\sim{\rm N}(0,0.04)
  \end{array}
\end{equation}

\subsubsection{Economic MPC}
The optimal steady states are ${{\boldsymbol{x}}_{\rm{s}}} = [0.5096~536.7485]^{\rm T}$ and ${{\boldsymbol{u}}_{\rm{s}}} = [10~200000]^{\rm T}$. The controller gain obtained from sensitivity analysis is shown in Eq. (\ref{eq:5.3.5}), in which the elements are all 0. It means the optimal inputs will not change and their constraints are active as long as the process operates around the given steady state. It can be explained by that the inputs are economic decision variables in the optimization of EMPC.
\begin{equation}\label{eq:5.3.5}
  {{\boldsymbol{K}}^{{\rm{EMPC}}}} = \left[ {\begin{array}{*{20}{c}}
  0&0\\
  0&0
  \end{array}} \right]
\end{equation}

The mean and variance of each variable and the average performance indices are shown in Table \ref{tab:5.3.1} and Table \ref{tab:5.3.2}, respectively. They illustrate that the proposed sensitivity-based dynamic performance assessment approach works well in the EMPC controlled system.

\begin{table}
    \caption{Mean and variance of each variable of EMPC in Case 3}\label{tab:5.3.1}
    \centering
    \begin{tabular}{llllllll}
      \hline
               &            &         $x_1$ &         $x_2$ &         $y_1$ &         $y_2$ &         $u_1$ &         $u_2$ \\
      \hline
           \multirow{2}{*}{$\mu$} &        Sim & 5.1004e-01 & 5.3662e+02 & 5.1006e-01 & 5.3662e+02 &     1.e+01 &     2.e+05 \\

               &        Cal & 5.0955e-01 & 5.3675e+02 & 5.0955e-01 & 5.3675e+02 &     1.e+01 &     2.e+05 \\
           \multirow{2}{*}{$\sigma^2$} &        Sim & 3.7723e-05 & 9.0709e-01 & 7.6350e-05 & 9.4428e-01 &          0 &          0 \\

               &        Cal & 3.6951e-05 & 9.3944e-01 & 7.6951e-05 & 9.7944e-01 &          0 &          0 \\
      \hline
    \end{tabular}
\end{table}

\begin{table}
    \caption{Calculated and simulated performance indices of EMPC in Case 3}\label{tab:5.3.2}
    \centering
    \begin{tabular}{lllll}
      \hline
                &  & 3$\boldsymbol\sigma$ zone & 4$\boldsymbol\sigma$ zone & 5$\boldsymbol\sigma$ zone \\
      \hline
            \multirow{2}{*}{$\bar J_{\rm{ep}}$} &        Cal & -5.0370e+07 & -5.0631e+07 & -5.0638e+07 \\

               &        Sim &  & -5.0639e+07 &  \\
            \multirow{2}{*}{$\bar J_{\rm{tp}}$} &        Cal & 2.9194e-04 & 2.9942e-04 & 2.9732e-04 \\

               &        Sim &  & 3.0030e-04 &  \\
      \hline
    \end{tabular}
\end{table}

\subsubsection{Tracking MPC}

In this case, the controller gain is varying with respect to the positions of states because ${\boldsymbol u}_{\rm s}$ locates on the boundaries. When the optimal $\boldsymbol u$ needs acting toward the boundaries, the actual inputs will be restricted inside the boundaries. On the contrary, there is no problem. Therefore, we calculate the controller gains with active constraints ${\boldsymbol {I_u}}=\boldsymbol I$ and inactive constraints ${\boldsymbol {I_u}}=\boldsymbol 0$, respectively:
\begin{equation}\label{eq:5.3.6}
  {\boldsymbol{K}}_{{{\boldsymbol{I}}_{\boldsymbol{u}}} = {\boldsymbol{I}}}^{{\rm{TMPC}}} = \left[ {\begin{array}{*{20}{c}}
  0&0\\
  0&0
  \end{array}} \right]
\end{equation}
\begin{equation}\label{eq:5.3.7}
  {\boldsymbol{K}}_{{{\boldsymbol{I}}_{\boldsymbol{u}}} = {\boldsymbol{0}}}^{{\rm{TMPC}}} = \left[ {\begin{array}{*{20}{c}}
  -2.0092e\!+\!00 & 5.6354e\!-\!02\\
  -2.1417e\!+\!04 & -5.7478e\!+\!02
  \end{array}} \right]
\end{equation}

\begin{figure}
    \centering
    \subfigure[State trajectories]{
      \label{subfig:5.3.3a} 
      \includegraphics[width=0.475\textwidth,angle=0]{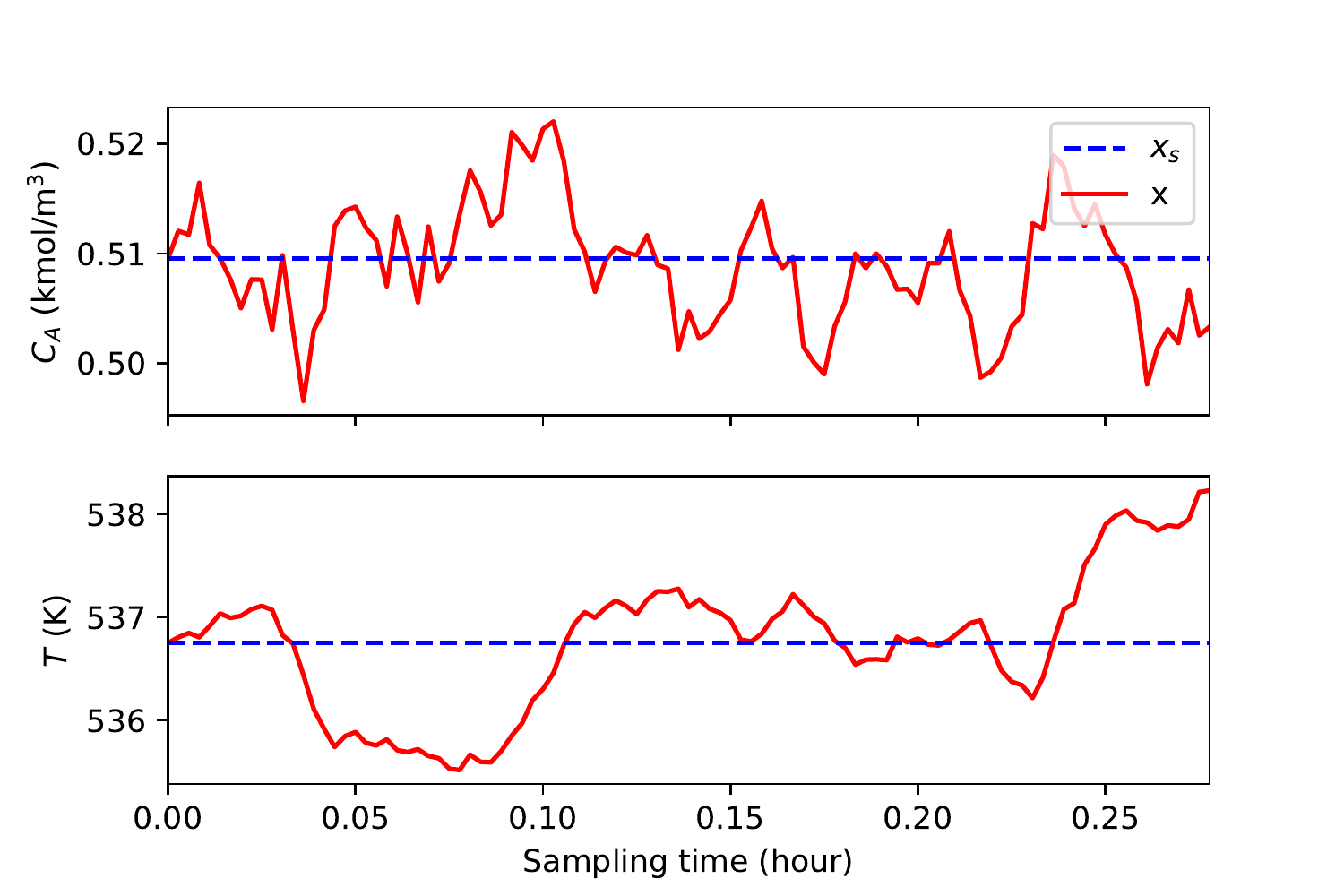}}
    \hspace{0in}
    \subfigure[Input trajectories]{
      \label{subfig:5.3.3b} 
      \includegraphics[width=0.475\textwidth,angle=0]{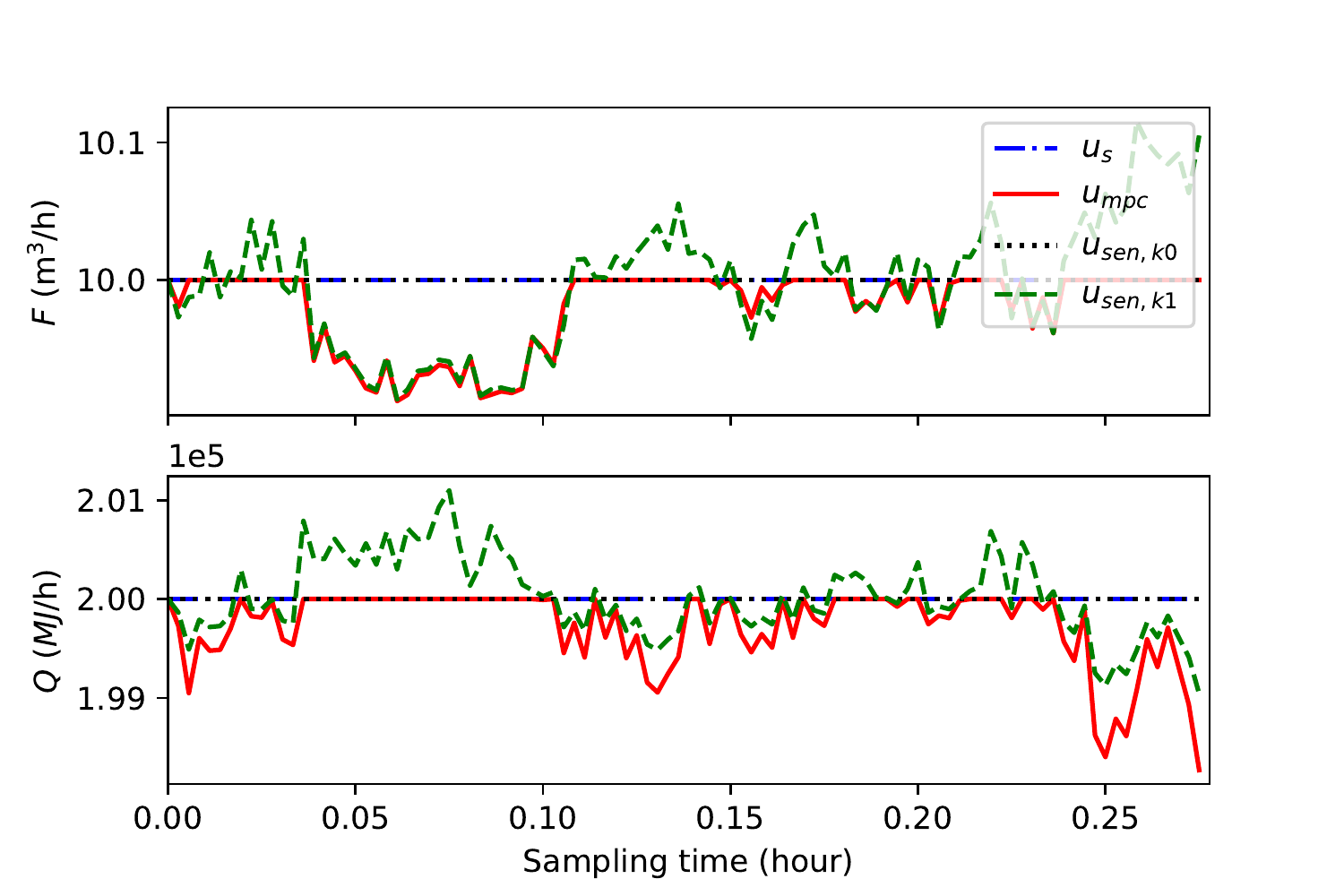}}
    \caption{State and input trajectories of tracking MPC in Case 3}\label{fig:5.3.3}
\end{figure}

It is impossible to know exactly the controller gain in this case, so here we just discuss and compare the actual actions of tracking MPC with the predicted results using sensitivity analysis. Figure \ref{fig:5.3.3} shows the trajectories of states and inputs, in which \textquotedblleft{$\rm u_{sen,k0}$}\textquotedblright\ denotes the input trajectory calculated using Eq. (\ref{eq:5.3.6}) and \textquotedblleft{$\rm u_{sen,k1}$}\textquotedblright\ denotes the results from Eq. (\ref{eq:5.3.7}). The possible trajectory and tendency of ${\boldsymbol u}_{\rm mpc}$ can be roughly predicted based on the results of sensitivity analysis. Some conclusions about the ideal economic performance, possible tendency of $\boldsymbol u$, and approximate economic loss can be roughly analyzed. It may be helpful for the engineers to understand the actual actions of inputs in the scenario. Actually, the situation that $\boldsymbol u$ locates on its boundaries should be avoided in tracking MPC controlled systems in consideration of the deficiency of control freedom. If target moving is applied, Case 3 will convert to be a ``Case 1'' problem and can be analyzed using the discussion in Section \ref{sec:5.1}.

\subsection{Case 4: Other conditions}\label{sec:5.4}
One of the largest differences between tracking MPC and EMPC is that the former is one kind passive target tracking which may lead to offset and other tracking problems, while the latter is a dynamic autonomous economy searching which will not result in offset problem for the overlook of given targets. Therefore, the proposed sensitivity-based dynamic performance assessment method works well with all EMPC controlled systems as long as the process and measurement noise is not too large.

\begin{figure}
    \centering
    \includegraphics[width=0.8\textwidth,angle=0]{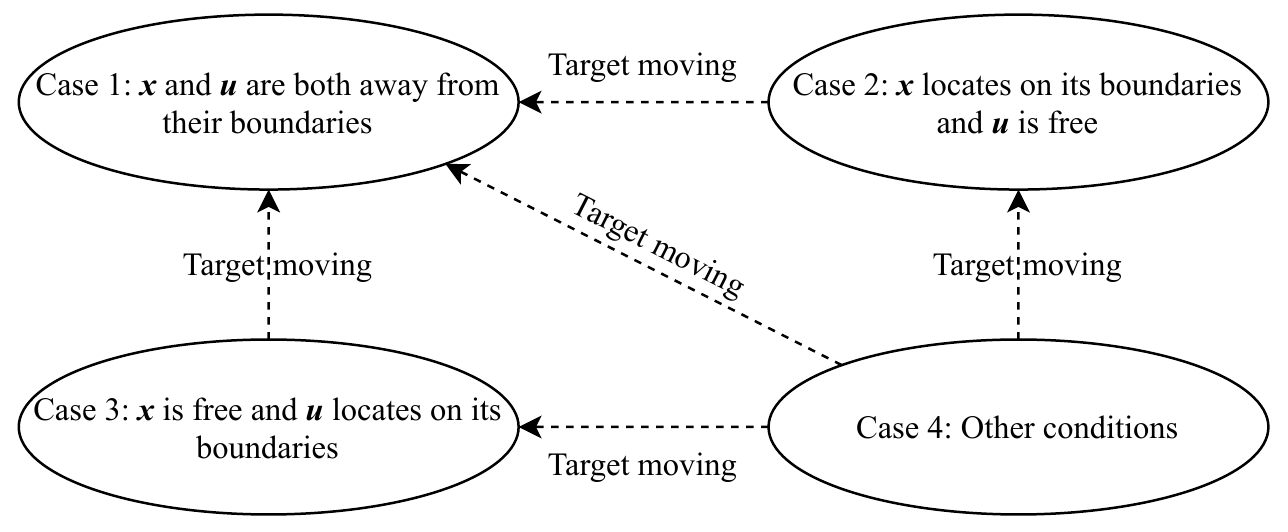}
    \caption{Different conditions of tracking MPC}\label{fig:5.4.1}
\end{figure}

However in tracking MPC controlled systems, different conditions need to be distinguished to make the analysis more reliable. Figure \ref{fig:5.4.1} gives a complete view on all possible conditions in tracking MPC. Case 1 - Case 3 have been discussed in detail in this work. Case 4, \textquotedblleft{Mixed active constraints}\textquotedblright, denotes the condition that some states and inputs are away from their boundaries and some locate on their boundaries. The case can be converted to be a new case problem discussed above through target moving.

\section{Conclusion}\label{sec:6}
In this work, a sensitivity-based dynamic performance assessment method is proposed to analyze the tracking and economic performance of tracking MPC and EMPC in the presence of constraints and Gaussian noise. It can provide effective guidance on which controller can achieve better performance for a given system and is very helpful for controller design. Extensive experiments on a CSTR process demonstrate the effectiveness of the proposed approach. In the future work, more concentrations will be focused on the dynamic performance assessment of MPC with colored noise and disturbance.



\end{document}